\documentclass[twocolumn]{aa}
\usepackage{natbib}
\bibpunct{(}{)}{;}{a}{}{,}
\usepackage{psfig}
\usepackage{epsfig,afterpage} 
\usepackage{graphicx}
\usepackage{pifont}
\usepackage{amssymb}
\usepackage{aalongtable}
\usepackage{array}
\usepackage{amsmath}
\usepackage{rotating}
\usepackage{multirow}
\usepackage{lscape}

\begin{document}

\title {
        Abundances of four open  clusters from  solar stars.
\thanks{Observations collected at the ESO VLT. 
Table  1 is only available   in electronic form at
the CDS via anonymous ftp to cdsarc.u-strasbg.fr (130.79.128.5) or via
http://cdsweb.u-strasbg.fr/cgi-bin/qcat?J/A+A/}} 

\subtitle{ }

\author{ 
        G. Pace \inst{1}$^,$\inst{2}, L. Pasquini \inst{3}, 
        \and P. Fran\c cois \inst{4}}

\offprints{G. Pace, \email gpace@aries.ernet.in}

\institute{
            Centro de Astrofisica, Universidade do Porto, 
            Rua das Estrellas, 4150--762 Porto, Portugal    \\  
            \and
            Aryabhatta Research Institute of Observational Sciences, 
            Manora Peak, Nainital, 263129 Uttaranchal, India    \\  
            \and
            European Southern Observatory, Karl Schwarzschildstr. 2,
            Garching bei M\"{u}nchen, Germany \\ 
            \and
            Observatoire de  Paris, 64  Avenue de l'Observatoire, 75014
            Paris, France \\}


          \abstract  {}{We  present   the  abundance  measurements  of
            several elements (Fe,  Ca, Na, Ni, Ti, Al,  Cr, Si) for 20
            solar--type   stars  belonging   to  four   Galactic  open
            clusters: NGC~3680,  IC~4651, Praesepe, and  M~67.  Oxygen
            abundances  were in  addition measured  for most  stars in
            each cluster  apart from IC~4651.   For NGC~3680, accurate
            abundance  determinations  using high--resolution  spectra
            covering  a large  spectral  domain are  computed for  the
            first  time.}   {  We  used  UVES  high--resolution,  high
            signal--to--noise  (S/N)  ratio  spectra and  performed  a
            differential  analysis   with  respect  to   the  sun,  by
            measuring equivalent widths  and assuming LTE.}  {The most
            surprising   result  is   a  measurement   of  significant
            supersolar         metallicity         for        Praesepe
            ([Fe/H]=0.27$\pm$0.10).   As for  the  other clusters,  we
            confirm    a    supersolar    metallicity   for    IC~4651
            ([Fe/H]=0.12$\pm$0.05),  a   solar  metallicity  for  M~67
            ([Fe/H]=0.03$\pm$0.04)  and a slight  subsolar metallicity
            for  NGC~3680 ([Fe/H]=-0.04$\pm$0.03).   We find  that the
            abundance ratios  of almost  all elements are  solar, with
            the notable exception of  oxygen in NGC~3680 and Praesepe,
            supersolar in the former cluster ([O/Fe]=0.2$\pm$0.05) and
            as low as [O/Fe]=-0.4$\pm$0.1 in the latter.  Observations
            of  several  objects per  cluster  is  required to  obtain
            robust  results,  especially  for  those elements  with  a
            limited number of suitable lines. }{}


\authorrunning{Pace et al.\ }

\titlerunning{Abundance in solar--type stars in four open clusters.}

\maketitle

\section{Introduction}

Understanding the  chemical evolution of the Galaxy  requires both the
development of theoretical models  and comparison of their predictions
with observational results. Several different models have been created
of  increasingly more  complex and  realistic scenarios,  for instance
including two infall episodes that formed the halo-thick disk and thin
disk   \citep{cmg97,cmg01,pat04}.    Observationally,   new-generation
telescopes such as VLT and  Keck and the availability of multi--object
spectroscopy  have  enabled  a  significant  amount  of  high  quality
spectroscopic

A long-standing question concerns  the existence and evolution  of the
chemical  abundance gradient   in  the Galactic  disk.   Galactic open
clusters are probably the best tool for  understanding whether and how
the gradient  slope changes with time because  they have formed at all
epochs.  Their distances can be measured  more accurately and are less
affected by observational biases than other classes of objects.

\cite{dl05}   identified the   birthplace of  612   open  clusters and
determined the spiral pattern  rotation  speed of the Galactic   disk.
The possibility of tracing open clusters back  to where they formed is
interesting because, coupled with  metallicity measurements,  it allow
us to reconstruct  the chemical distribution of  the Galactic disk  in
the past.  \cite{dl05} based their work on large photometric databases
in the visible.  They  achieve important conclusions on the  structure
and   dynamics of  the spiral   arms,  but to   estimate precisely the
deviation of open cluster  motions  from circular orbits, and  improve
the scale of  Galactocentric distances at  the place of birth, a  more
complete  and precise database   of age, distance,  proper  motion and
radial velocity    determinations   is required.         Unprecedented
opportunities    are provided by   deep   photometric  surveys in  the
infrared, such as UKIDSS \citep{ukidss} and VISTA \citep{vista}. These
surveys   can be used to  complete  deep, precise and homogeneous open
cluster  databases of location in the   3-D space, age, and reddening.
The calibration  of photometric metallicity  indicators could  be used
for clusters not observable with high--resolution spectroscopy.

Open clusters have the added advantage of providing a sample of coeval
stars that formed from  the same material,  which means, in particular
for main--sequence stars,  that they should  have the same atmospheric
chemical   composition.  As a result,   the chemical composition of an
open cluster can   be studied by  several  stellar spectra.  With  all
these  advantages,  open  clusters are   ideal   objects to probe  the
chemical evolution of the Galactic disk.  A large dataset, spanning as
wide a range of Galactocentric distances and ages,  as possible, is of
course required.

Finally, clusters are  at the basis  of  our understanding of  stellar
evolution,  and  their colour--magnitude diagrams, and well-determined
abundances  can be   used to test  stellar   evolution models directly
\citep[see e.g.][]{naa97}.

To  date, accurate chemical composition data  have been determined for
only a few open clusters using high  resolution spectra of a number of
primarily giant stars  \citep{gratton00}.   The situation is  evolving
rapidly due  to coordinated  efforts   using facilities available   at
8--M--class  telescopes   \citep[][]{largeprog,    BOCCE}.  This  work
combines with   other  efforts  attempting  to   establish  a   robust
open-cluster  metallicity  scale.  Random  and  systematic errors have
given rise  to  -- sometimes dramatic  --  discrepancies  in abundance
determinations  of a  given open  cluster  as  computed from different
groups.  We confirm  that this is no longer   the case when  comparing
independent      metallicity   measurements    from  high   resolution
spectroscopy.

\section{Observations and data reduction}
\label{obs}

The data  used  were primarily collected   to study the  chromospheric
activity evolution   of solar--type stars \citep{paper1}.   The sample
includes two stars  in NGC~3680, five in  IC~4651, seven  in Praesepe,
and six in M~67.  The  targets were chosen  to be main sequence stars,
and to have a  high probability of being  members of the  clusters and
not known to  be binaries at the  time of observations.  Our selection
was completed using the reference works  of \cite{naa96} for NGC~3680,
\cite{man02}  for  IC 4651,  and  several sources for  M~67, including
\cite{lmmd92} for the binary  determination in this cluster.  From the
sample originally  collected and used in  \cite{paper1}, one M~67 star
was excluded because it turned  out to be a binary \citep{lastsofias}.
Our stars have colours in the  range 0.51 $<$(B$-$V)$_0 <$ 0.72, which
includes the solar colour evaluated  around B$-$V = 0.65 \citep[see e.
g.][]{pb08}.  We note that the  paucity of stars observed in  NGC~3680
is due to  the fact that,  as shown by  \cite{naa97}, this cluster has
few  single G stars  members, since most  of its low mass members have
being dispersed during its lifetime.

Star names are taken from \cite{eggen69}  for NGC~3680, \cite{AMC} and
\cite{eggen71}  for IC~4651, \cite{sand}  for  M~67, and \cite{kw} for
Praesepe.

The spectra were obtained during  the ESO observing run 66.D-0457 with
the  UVES  spectrograph   at  the  focus  of  Kueyen   2  of  the  VLT
\citep{uves}.  While we used the blue part of the spectra to determine
the level  of chromospheric activity,  we did also record  red spectra
simultaneously for the range between  480 and 680 nm. Given the higher
flux of stars in this range, we could reduce the width of the UVES red
slit to 0.4  arcseconds, to achieve a resolution  of R=100\,000 at red
wavelengths,  while maintaining a  high S/N  ratio in  this wavelength
range.  With  some variation from  star to star and  between different
regions of the spectra, the  S/N ratio/pixel is about 130 for Praesepe
stars and 80  for all other stars.  The spectra  were reduced with the
UVES pipeline \citep{pipeline}, and then analysed using both MIDAS and
IDL routines

\section{Abundance analysis}
\label{abundanceanalysis}

Abundance measurements were completed  by measuring  equivalent widths
(EWs) and  using  OSMARCS models  \citep{osmarcs},  in a standard--LTE
analysis.  We used the line  list  from \cite{lastsofias}, from  which
suitable  lines in  the EW--range 5--140   m{\AA}  were selected.   EW
measurements were performed by  using an IDL  program developed by one
of the authors (G.  P.), which operates  semi-automatically to allow a
visual control of the fit of each selected line.

We  first   completed  the  chemical analysis    of  the Sun    for EW
measurements  determined   by \cite{lastsofias}, and   the well--known
values  of temperature  and gravity  \citep[see  e.g.][]{sunpar}.  The
solar EW  measurements used were  obtained for  an UVES  spectrum of a
resolution R=45\,000.  A comparison with  EW measurements for the UVES
archive solar  spectrum acquired  for the  same  configuration as  our
sample spectra, indicated that variations  in resolution power between
R=45\,000 and R=100\,000 do not have any detectable effect on the EWs,
and we therefore    preferred to employ   published  measurements.  We
attempted    to  describe  the data     using several values   for the
microturbulence and  eventually   chose the value that   provided  the
flattest trend in the EW  versus abundance diagram, namely 1.1 km/sec.
Our  results are  available   at  CDS  in  electronic form  (Table  1)
containing the following information.  Column  1 lists the wavelength,
Col.   2 indicates the  element  chemical  symbol, Col.  3  ionization
stage, Col.  4 indicates the EW  measurement, and Col.  5 provides the
corresponding chemical abundance.   While the correlation  coefficient
between [Fe/H] and EW is close to  0, namely 0.02, that between [Fe/H]
and the excitation potential   ($\chi$) is -0.42, and  the  difference
between Fe {\sc I} and  Fe{\sc II} is  approximately -0.05 dex.  These
numbers differ significantly  from  zero.  Small inaccuracies in   the
model   that   reproduce the  atmospheric  temperature stratification,
rather than the limits of the LTE assumption, could be responsible for
the   slope in  the [Fe/H]  versus    $\chi$ diagram and    for a poor
ionization equilibrium,  since lines of different excitation potential
tend to form in different layers of the solar atmosphere \citep{gs99}.

\addtocounter{table}{1}

By performing a differential analysis line by line with respect to the
Sun, we expect to reduce significantly the spurious trends and restore
the ionisation  equilibrium  when evaluating   the stellar parameters,
provided that the structure of the stellar photosphere does not differ
significantly from  solar, which is  a reasonable assumption since our
targets are solar--type stars.

\begin{table}
\begin{center}
\begin{tabular}{c c c c c}
X &A(X)$_{our}$&A(X)$_{litt}$&Number&$\sigma$\\
&&&of lines&\\
\hline

Fe{\sc I} & 7.52 & 7.50  &66&0.03\\
Fe{\sc II}& 7.57 & 7.50  &11&0.03\\
Na{\sc I} & 6.37 & 6.33  & 3&0.11\\
Al{\sc I} & 6.47 & 6.47  & 2&0.02\\
Si{\sc I} & 7.56 & 7.55  & 9&0.03\\
Ca{\sc I} & 6.36 & 6.36  &11&0.03\\
Ti{\sc I} & 4.97 & 5.02  &11&0.02\\
Cr{\sc I} & 5.65 & 5.64  & 6&0.06\\
Ni{\sc I} & 6.25 & 6.25  &23&0.04\\
\hline
\end{tabular}
\end{center}
\caption
{ Comparison between our solar abundances (second Col.) and those of
Grevesse \&  Sauval  (2000) (third Col.).   For  iron, the abundance
value obtained using ionised lines is also given.  The number of lines
and the  standard deviation of  the measurements  for each species are
also  reported  in the fourth and  in  the fifth  Col. respectively.
When only three or two lines are  measured (sodium and aluminum), half
of the difference between the  maximum and  minimum value is  reported
instead of the standard deviation.}  
\label{solarres}

\end{table}

For stars     studied, atmospheric   parameters that   reproduce  most
effectively excitation and ionisation  equilibria were selected from a
broad grid of guess values.

The temperature estimates about which the grids were constructed, were
computed first for existing photometry, in particular V-H, V-J and V-K
colours.  V--magnitudes were taken  from the reference papers in Sect.
\ref{obs}   for   IC~4651, from   \cite{naa97}   for   NGC~3680,  from
\cite{mmj93}  for  M~67, and  from \cite{js91} for  Praesepe, with the
exception of KW~368 and  KW~208, whose V--magnitudes are taken instead
from  \cite{jc83} and \cite{j52} respectively.  We  then adopted J, H,
and K magnitudes from the 2~MASS catalogue \citep{2mass}.  Each of the
V-H, V-J, and   V-K  colours  were   transformed into  a   temperature
estimation  using a calibration defined  by \cite{rm05}, and the three
values were averaged.

The   comparison between  photometric  and spectroscopic  temperatures
indicated     that   the   colour--colour   and    colour--temperature
transformations  could be affected  by  systematic error, and that the
colour excess adopted  for IC~4651 could  be underestimated (see Sect.
\ref{spph}).

As for gravity, we used the  values expected for a dwarf according to:
1) the  previously determined photometric temperature;  2) the stellar
ages, i.e.  those of the parent cluster given in \cite{cc94} and, only
for  Praesepe, in  the WEBDA  database; and  3) the  solar metallicity
isochrones of \cite{leo00}. Microturbulence was allowed to vary in the
range  from  $\approx  0.7$  to $\approx  1.7$  km/sec.   Temperature,
gravity and  microturbulence velocity were set equal  those values for
which  no  significant  trend  in  the computed  iron  abundances  was
present, neither as a function of the excitation potential of the line
nor as a function of its EW, and for which the iron abundances derived
using the Fe {\sc I} and Fe {\sc II} lines provided similar results to
within  the margins  of  error.  The  correlation  coefficients of  EW
versus  abundance   and  $\chi$   versus  abundance  are   in  general
$\approx$~0.1 or smaller, always smaller  than 0.3, and Fe {\sc I} and
Fe{\sc II}  provide consistent measurements that agree  to within 0.03
dex.  Praesepe stars  are an exception because we  were unable to find
parameters  as  good  as  those  obtained  for  the  other  stars  and
consistent  with the  photometric  data, and  in  some cases  accepted
correlation coefficients as high as  0.5.  The explanation for this is
that we  have used the  same line list  for all the  sample, therefore
Praesepe stars, which are more metal rich and have an EW range that is
shifted towards larger values, have  a higher percentage of lines that
are close to saturation and fewer weak lines, hence a less constrained
microturbulence.  Furthermore, microturbulence is significantly higher
in Praesepe stars than in the others.  Differences in the structure of
the  photosphere, due  to  higher metallicity  and rotation  velocity,
could produce a  higher microturbulence.  In any case,  as we see, our
measurement  of a  significantly higher  chemical abundance  for these
stars is robust.

For each    tentative   combination  of  temperature,   gravity,   and
microturbulence values of the grid,  the metallicity was initially set
to solar, and then iteratively  substituted with the average abundance
corresponding to the Fe {\sc I} lines, until convergence was achieved.
Furthermore, for each  measured line,  the abundances were  calculated
differentially with respect  to  the sun.  This,  in addition   to the
afore--mentioned more  robust parameter   determination, allows us  to
compensate for errors in the $\log~gf$ values \citep{langer98}.

The steps  in the grid values are  20 K for temperature,  0.02 dex for
$\log G$,  and 0.02 km/sec  for $\xi$.  We  used between 54 and  64 Fe
{\sc I}  and between 9 and  11 Fe {\sc  II} lines for each  star.  For
only three  Praesepe stars, no  higher than 7  or 8 Fe {\sc  II} lines
could be measured reliably, and for  one star, no more than 42 Fe {\sc
  I} lines.

As expected,  we   found a  correlation between   microturbulence  and
temperature obtained  as described above.   Praesepe stars, as already
discussed, have a   significantly  higher microturbulence value;    we
therefore  calibrated      two  different    linear   T--versus--$\xi$
relationship, one for Praesepe, and one for the remaining stars.

We  fine--tuned the parameters, to  ensure that  $\xi$ was as close as
possible to the value expected from the calibrations, and the
gravity as consistent as  possible with the spectroscopic temperature,
not anymore with the photometric one.

We  calibrated the T--versus--$\xi$ relationship for the second time,
with the finally  adopted  parameters, not using  Praesepe  stars. The
result is:

\begin{math} 
\xi=-5.42+6.30\cdot\frac{T_{eff}}{T_{\odot}},\  \ \  \ \sigma=0.06  \\
\end{math}
here $\sigma$  represents the standard deviation   of datapoints about
the fit. For the remainder  of the paper, we  refer to $\xi_{fit}$
as  the  microturbulence values derived  from  the fit above.

Abundances were   finally  recomputed   by assuming the    fine--tuned
parameters.

In  cluster   stars, we  measured  the  abundance  of oxygen, which is
treated separately, iron, calcium, aluminum, sodium, nickel, titanium,
chromium, and  silicon. As for  iron lines, we used solar measurements
as reference  values for all of our  [X/H] estimates of cluster stars,
computing for all elements the  difference between the stellar and the
solar  value for each line measured.   

Oxygen abundances,  instead,  were derived  from  measured EWs of  the
O~{\sc i} 6300.30~\AA~forbidden line employing the same method used by
\cite{lastsofias}.  Specifically,  we employed  MOOG  \citep[][version
2002]{sneden} and Kurucz model atmospheres \citep{kur93}.  We used the
driver  {\it blend},   which   allowed us  to  take  into  account the
contribution  of    the blending  Ni~{\sc i}  6300.34   feature to the
measured $EW$s of the  6300.3~\AA~feature.  As input  to MOOG, we used
the results of our analysis for stellar parameters and iron and nickel
abundances.  For  the oxygen and  nickel lines we employed $gf$-values
equal $\log~gf=-9.717$ and $\log~gf=-2.11$, respectively,    following
\cite{ap01}     and   \cite{joha03}      \citep[see][for    additional
details]{lastsofias}.

Because of  the cluster radial velocities,  we did not need to correct
for the presence of  telluric lines for  NGC~3680, Praesepe, and M~67;
on  the other  hand, telluric  lines severely  affected the spectra of
IC~4651 stars,  and  we were unable  to  correct for  this due to  the
unavailability of suitable standard stars. Finally, for three stars in
Praesepe and one in M~67 we were unable to measure the O~{\sc i} line.

\addtocounter{table}{1}

\section{The results}
\label{results}

The results for the Sun are  summarised in Table \ref{solarres}, those
for the cluster stars in Table \ref{metmeas}.

The   results shown  in Table   \ref{solarres}  were obtained for  the
(T=5780, G=4.40, $\xi$= 1.1) model.  In Col. 1 we indicate the element
and ionization  stage. In Col.  2 we wrote  our results, and in Col. 3
the values from \cite{solarmet} are given as comparison.

For  each star,  the mean  abundance values  of all  elements studied,
apart from  oxygen are  given in Table  \ref{metmeas}, along  with the
error calculated by the  analysis described in Sect. \ref{sec_errors}.
The  standard  deviation  corresponding  to the  measurements  of  the
different lines, and the number of lines measured for each element are
also given (when no more than  three lines could be measured, half the
difference between the  highest and lowest values is  given instead of
the standard deviation).  All  values refer to neutral--element lines.
The stars are  grouped by cluster, and for  each cluster one abundance
is obtained  by averaging  the abundance of  all stars.   The relative
standard  deviation is  also computed,  apart from  for  NGC~3680, for
which  only  2  stars  were  observed.   In this  case,  half  of  the
difference  between  the  two  stellar  abundances  is  given.   These
standard deviations indicate the  robustness of our analysis.  Only in
four cases (aluminum in Praesepe and M~67 and titanium and chromium in
M~67)  the  cluster abundance  dispersion  is  larger  than 0.07  dex.
Furthermore, we note that the  dispersion in the stellar abundances of
M~67 is significantly enhanced by Sanders 1287, which always shows the
lowest chemical content.  Its case is worth further investigation.  On
the other hand, the maximum difference for stars belonging to the same
cluster can easily  exceed 0.1 dex, which implies  that caution should
be taken when deriving conclusions based on a small number of objects.
To  represent  errors  in  the  final  values,  we  use  the  standard
deviations of the stellar abundance measurements in each cluster.

Since for Praesepe,  in the parameter determination, higher EW--[Fe/H]
and $\chi$--[Fe/H] correlations  are  sometimes found,  we assume  the
more  conservative  average error    estimate  for its  single   stars
originating in the  analysis described in Sect. \ref{sec_errors} of 0.10
dex. By   doing  so we  account for  the  possibility  of a systematic
component in the parameter determination of Praesepe stars.

\subsection{Errors}
\label{sec_errors}
There  are   three sources of   measurement   errors in  our abundance
analysis:

\begin{list}{$\cdot$}{\itemindent -15pt   \itemsep  -3pt }  
\item error in EW measurements;
\item error in atmospheric parameters;
\item error in $\log gf$ values.
\end{list}

The last  item is the  least significant,  since  it should be  almost
eliminated  when  subtracting the  solar  abundances  from the stellar
ones.   On the other  hand, EW measurements  and atmospheric parameter
determinations are  affected   by  errors that   are  not  necessarily
negligible for in the stellar and the solar estimations.

\subsubsection{Errors in equivalent width measurements and $\log gf$}
\label{errew}

Errors in  EW measurements are due   mostly to the uncertainty  in the
choice of  the   continuum; we should   therefore  underestimate these
significantly by using the  Cayrel formula \citep{cayrel88}.  They are
by far  the  most significant  contributors to the   dispersion in the
abundance   measurements for  different  lines  of  the same  element.
Divided by the square root of  the number of  lines, the dispersion in
the abundance measurements has to be quadratically  added to the other
contributions to compute the global error.

\subsubsection{Errors in the parameter atmosphere}
\label{paramerr}

To evaluate the uncertainty in the parameters, we used the differences
between    those  derived in  our  spectral    analysis, $T_{spec},\ \
G_{spec}$, and $\xi_{spec}$;    and the computed  ones, $T_{phot},\  \
G_{phot}$,  and $\xi_{fit}$.   The  evaluations of   $T_{phot}$ and  $
G_{phot}$ and the linear regression used  to compute $\xi_{fit}$, were
discussed in  Sect.  \ref{abundanceanalysis}.  The uncertainty  in the
gravity was  calculated  to be the   quadratic average (equal  to  the
standard deviation) of $G_{spec}-G_{phot}$.  We adopted the rms of the
$\xi_{spec} - \xi_{fit}$ differences to  represent the uncertainty  in
$\xi$.    We  remind the    reader  that the   linear fit   indicating
$\xi_{fit}$ as a function   of temperature was computed without  using
Praesepe  stars.  The  uncertainty in  the   microturbulence, which is
higher, was computed separately using the same formula.

For the temperature,  we used a  slightly modified approach; if we had
proceeded in the same way, we would have overestimated the uncertainty
in temperature and  been   unable to  determine   the errors  in   the
reddening,      which      may  contribute    to     the   differences
$T_{spec}-T_{phot}$.    We  therefore  computed   the   mean value  of
$T_{spec}-T_{phot}$ for each cluster, and assumed  that the mean value
was  due  to an  incorrect evaluation  of  the reddening.   Only after
subtracting this value from each term did we proceed in evaluating the
standard deviation of $T_{spec}-T_{phot}$.

The final values  for the uncertainty in the  parameters are:  

\noindent $\Delta  T=110$ K,  $\Delta G=0.07$  log(gr$\cdot$cm$\cdot$sec$^{-2}$),
$\Delta \xi=0.06$ km$\cdot$sec$^{-1}$, 

\noindent $\Delta \xi=0.18$ km$\cdot$sec$^{-1}$ for Praesepe

\subsubsection{Errors in the final abundance values.}

To evaluate the uncertainty in the final abundance due to the error in
a particular atmospheric parameter ($T_{eff}$,  $\log~G$ or $\xi$), we
repeated the entire chemical analysis twice: in the first analysis, we
increased the  value of the  parameter,  and in  the second  analysis,
decreased its value, by an amount equal  to the error in the parameter
computed as  described   in Sect.   \ref{paramerr},  while  all  other
parameters were kept constant.   We then completed a similar procedure
to determine the  errors in  the  EW measurements  and $\log~gf$  (see
Sect.   \ref{errew})  and   we  finally added  quadratically   all the
contributions.

However, as discussed above, we propose that a more robust estimate of
our  final  errors in the abundances  are  represented by the standard
deviation values.    The  abundance  measurements of   different stars
belonging to    the  same  cluster  actually    represent  independent
evaluations of the same quantity, which is the cluster abundance.  The
error analysis in this Sect. is supposed to determine which parameters
influence  the final determination    most significantly, and  confirm
whether our method for evaluating such parameters is robust.

We find that  the uncertainty in  the temperature  dominates the final
error,   since the other contributions are   in most cases marginal or
even negligible, and that  the assumed uncertainty  of 110 K, which is
due   to the different   methods  of measurement  for photometric  and
spectroscopic data, appears to slightly overestimate the errors.

As for the  EW, although for a single  line, the measurement error can
reach up to $\sim 20 \%$, having at our disposal many lines means that
its contribution  to the total  error is  far less important  than the
uncertainty in the temperature.

According to  these tests, silicon is  more insensitive than the other
elements to the  uncertainties.  It  is  remarkable as shown  in Table
\ref{metmeas}   the dispersion in   measurements  of  [Si/H]  for each
cluster is among  the smallest, and is  the  smallest dispersion value
for all elements for IC~4651 and  NGC~3680 despite the small number of
lines  used.  This  again points towards    a good consistency in  our
analysis.

\thispagestyle{empty}
\begin{table*}[ht] 
\begin{center}
\begin{footnotesize} 
\begin{tabular}{l c c c c c c c c c c c c c c }

CLUSTER%
&$|\frac{\partial [Fe/H]}{\partial \xi}\cdot \Delta \xi|$ %
&$|\frac{\partial [Fe/H]}{\partial   T}\cdot \Delta   T|$ %
&$|\frac{\partial [Fe/H]}{\partial   G}\cdot \Delta   G|$ %
&$|\frac{\partial [Fe/H]}{\partial  EW}\cdot \Delta  EW|$ \\
\noalign{\smallskip}					      
\hline
     NGC 3680 &  0.01&  0.09&   0.01& $<$0.01 \\
      IC 4651 &  0.02&  0.09&   0.01& $<$0.01 \\
     PRAESEPE &  0.06&  0.09&   0.01& $<$0.01 \\
         M 67 &  0.02&  0.09&   0.01& $<$0.01 \\   
\hline
&$|\frac{\partial [Na/H]}{\partial \xi}\cdot \Delta \xi|$ %
&$|\frac{\partial [Na/H]}{\partial   T}\cdot \Delta   T|$ %
&$|\frac{\partial [Na/H]}{\partial   G}\cdot \Delta   G|$ %
&$|\frac{\partial [Na/H]}{\partial  EW}\cdot \Delta  EW|$ %
&$|\frac{\partial [Na/H]}{\partial met}\cdot \Delta met|$ \\
\hline
     NGC 3680 &  $<$0.01&  0.05&    0.01&    0.01&  0.01 \\
      IC 4651 &     0.01&  0.06&    0.01&    0.02&  0.01 \\
     PRAESEPE &     0.02&  0.06&    0.01&    0.02&  0.01 \\
         M 67 &     0.01&  0.06&    0.01&    0.01&  0.01 \\   
\hline
&$|\frac{\partial [Ni/H]}{\partial \xi}\cdot \Delta \xi|$ %
&$|\frac{\partial [Ni/H]}{\partial   T}\cdot \Delta   T|$ %
&$|\frac{\partial [Ni/H]}{\partial   G}\cdot \Delta   G|$ %
&$|\frac{\partial [Ni/H]}{\partial  EW}\cdot \Delta  EW|$ %
&$|\frac{\partial [Ni/H]}{\partial met}\cdot \Delta met|$ \\
\hline
     NGC 3680 &     0.01&  0.06&$<$0.01&   0.01&  0.01\\
      IC 4651 &     0.01&  0.06&$<$0.01&   0.01&  0.01\\          
     PRAESEPE &     0.03&  0.06&$<$0.01&   0.01&  0.01\\
         M 67 &     0.01&  0.06&$<$0.01&   0.01&  0.01\\   
\hline
&$|\frac{\partial [Si/H]}{\partial \xi}\cdot \Delta \xi|$ %
&$|\frac{\partial [Si/H]}{\partial   T}\cdot \Delta   T|$ %
&$|\frac{\partial [Si/H]}{\partial   G}\cdot \Delta   G|$ %
&$|\frac{\partial [Si/H]}{\partial  EW}\cdot \Delta  EW|$ %
&$|\frac{\partial [Si/H]}{\partial met}\cdot \Delta met|$ \\
\hline
     NGC 3680 & $<$ 0.01&  0.02&$<$0.01&   0.01&  0.01\\
      IC 4651 & $<$ 0.01&  0.02&$<$0.01&   0.01&  0.01\\          
     PRAESEPE &     0.01&  0.02&$<$0.01&   0.02&  0.02\\
         M 67 & $<$ 0.01&  0.02&$<$0.01&   0.01&  0.02\\   

\end{tabular}
\end{footnotesize}
\end{center}
\caption{Table of   the  typical   errors associated   with  abundance
measurements of iron  (about sixty Fe  {\sc I} lines),  sodium (two or
three lines), and nickel (about twenty lines).}
\label{taberr}
\end{table*}

\subsection{Oxygen}  

The measured EWs of the 6300.30~\AA~line, and  derived n(O) and [O/Fe]
values are   listed     in  Table \ref{tab:oxy}.   We    recall   from
\cite{lastsofias} that by using the same method  and log- $gf$ values,
we derive a solar oxygen abundance n(O)$_{\odot}$=8.66.

\begin{table*}
\begin{center}
\begin{tabular}{lccc} 

Star        &EW 6300.30 \AA~(mA)    &     n(O)             & [O/Fe]                   \\ 
\hline
AHTC 1009   & 7.0 $\pm 1$           &8.91 $\pm 0.09$       & $ 0.25 \pm 0.13$         \\
Eggen 70    & 4.7 $\pm 0.8$         &8.75 $\pm 0.07$       & $ 0.16 \pm 0.11$         \\
\hline											                    
NGC 3680    &                       &8.83 $\pm 0.08$       & $ 0.20 \pm 0.05$         \\                    
\hline											                    
KW 100      &  --                   &--                    & --                       \\
KW 208      &  --                   &--                    & --                       \\
KW 326      & 5.4 $\pm 0.8$         &8.51 $\pm 0.17$       & $-0.44 \pm 0.20 $        \\
KW 368      & 5.0 $\pm 0.8$         &8.40 $\pm 0.17$       & $-0.52 \pm 0.20 $        \\
KW 392      & 4.8 $\pm 0.5$         &8.69 $\pm 0.10$       & $-0.32 \pm 0.15$         \\
KW 418      & 4.0 $\pm 1$           &8.48 $\pm 0.20$       & $-0.42 \pm 0.22$         \\
KW 49       &  --                   &  --                  &    --                    \\
\hline											                    
PRAESEPE    &                       &8.52 $\pm 0.12$       & $-0.42 \pm 0.08$         \\                    
\hline											                    
Sanders 1048& 4.4 $\pm 0.6$         &8.49 $\pm 0.09$       & $-0.20 \pm 0.13$         \\
Sanders 1092& 4.5 $\pm 1.5$         &8.67$^{+0.19}_{-0.32}$& $-0.06 ^{+0.21}_{-0.33}$ \\
Sanders 1255& 5.3 $\pm 0.4$         &8.68 $\pm 0.06$       & $ 0.01 \pm 0.12$         \\
Sanders 1283&   --                  & --                                              \\
Sanders 1287& 4   $\pm 1  $         &8.58$^{+0.14}_{-0.19}$& $-0.04 ^{+0.17}_{-0.21}$ \\
Sanders 746 & 5.3 $\pm 0.7$         &8.65 $\pm 0.1$        & $-0.08 \pm 0.13        $ \\
\hline
M 67        &                       &8.61 $\pm 0.08$       & $-0.07 \pm 0.08$         \\ 
\hline
\end{tabular} 
\caption{
EWs  of the 6300.30~\AA~O~{\sc i}  line  derived oxygen abundances. In
both columns, the  global cluster errors for  the cluster  refer to the
standard deviations in the abundances originating in the single stars,
while reported errors in single  star abundance measurements come from
the  uncertainty in the EWs.  See the text for  the  errors due to the
parameter uncertainty.
}
\label{tab:oxy}
\end{center}
\end{table*}

The  listed  errors  in  n(O)  values   include the  contribution   of
uncertainties in the  measured $EW$s of  the forbidden line and errors
in Ni abundances (which  are, however, negligible).  The uncertainties
in stellar parameters, i.e.  $\Delta  T = \pm 110$~K,  $\Delta G = \pm
0.07$~dex, and $\Delta  \xi = \pm 0.06$~km/s,  correspond respectively
to the  uncertainties $\Delta  n(O)=-0.08/0.05$, $\pm 0.04$,  and $\pm
0.02$.  The uncertainty  of  $\pm 0.19$~km/s for   the microturbulence
velocity  in  Praesepe    stars,  corresponds,  instead,   to   $\Delta
n(O)=0.06/-0.07$.

Based  on  a single, faint   line, these measurements are  affected by
considerable uncertainty.   On the other,  hand  the average value for
each cluster has a standard deviation that is smaller than the typical
uncertainty in the [O/Fe] estimation of each single star.

\subsection{Comparison with published data}

In this Section, we provide  an overview of  the published data  about
chemical abundances in our target  clusters, and compare them with our
results.  The  content   of  the  Sect.     is summarised  in    Table
\ref{comptab},     in   which we  do    not   consider photometric and
low--resolution spectroscopy studies.  We add the data compilation for
Collinder~261, NGC~6253,  and  Berkeley~29, which  have  been  studied
twice, to compare results  from independent sets of chemical  analysis
and  verify  the overall reliability    of abundance measurements from
high--resolution spectroscopy.

\addtocounter{table}{1}

\subsubsection{NGC 3680}

For    this   cluster  \cite{at04}     obtained   a  measurement    of
[Fe/H]=-0.14$\pm$0.03, which was in  agreement with the measurement of
\citet{prp01} of [Fe/H]=-0.17$\pm0.12$.  The former analysis was based
for CCD photometry for the intermediate--band {\textit uvbyCaH$\beta$},
and the latter was instead based on a  small part of a high dispersion
spectrum of a single giant.

\cite{frieletal02} published  radial velocities and  metallicities for
39 clusters  older than the  Hyades.  They used spectra  of resolution
4~{\AA}~FWHM.  We studied two clusters  in their target list, M~67 and
NGC~3680,  for   which  they  measured  [Fe/H]=-0.15   and  -0.19  dex
respectively, with a standard error of 0.05 dex for both values.  They
studied 25  stars in M~67  and 7 in  NGC~3680.  The contrast  with our
result  is  significant,  but  both  our  measurements  and  those  of
\cite{frieletal02} point to a difference of about 0.05 dex between the
two  clusters,  NGC~3680  being  more metal  poor.   The  disagreement
between eech  of the afore--mentioned measurements  of the metallicity
in NGC~3680  and that of  [Fe/H]=+0.11 given by \cite{naa97}  based on
{\textit  uvby$\beta$} photometry,  cannot be  attributed  entirely to
measurement errors.

Our measurement   of  a  slightly metal--poor composition    ([Fe/H] =
-0.04$\pm0.03$)   is   the    first   to   our   knowledge based    on
high--resolution spectra of a wide wavelength range.


\subsubsection{IC 4651}

For   IC~4651 several   measurement  of   metallicity   are available.
\cite{przhcn} obtained their results using UVES spectra. They observed
twenty stars, including both dwarfs and giants.  A considerable effort
was made  to study stars  corresponding to the  region around the turn
off.   They measured an   iron  abundance of [Fe/H]=0.10$\pm$0.03,  in
excellent agreement with ours.   \cite{cbgt04} observed 4  clump stars
with the  high--resolution spectrograph FEROS at  the focus of the ESO
1.5--meter telescope,   and measured [Fe/H]=0.11$\pm0.01$.    In  this
case, the agreement in the measurement of iron abundance is excellent,
and it  is  remarkable that the  same  iron content  is found for both
giants and dwarfs.   \cite{man02} also measured [Fe/H]=0.1 dex,  using
{\textit  uvby$\beta$} data.  For   this cluster, the  measurements of
[Fe/H] obtained by independent studies appear to be in remarkably good
agreement.

As  far  as   other  elements  are  concerned,  we   can  compare  our
measurements with those of  \cite{przhcn}, who measured abundances for
several elements in common with  our study.  The measurements for main
sequence stars  are in good agreement  (within 0.03 dex)  with our own
for calcium, aluminum, and  nickel.  The discrepancy found between our
measurements of sodium abundances is  also within the margins of error
(0.06  dex, which  corresponds to  approximately 1$\sigma$).   For the
remaining  elements (silicon,  titanium and  chromium)  the difference
ranges from  0.09 to 0.11 dex.   We note that, while  we consider only
data for main--sequence stars, the values given in Table \ref{metcomp}
from \cite{przhcn}  correspond to  the overall cluster  abundance, and
the agreements between our measurements and theirs are poorer.  Within
a  margin  of  uncertainty of  2  $\sigma$,  all  the results  show  a
substantial solar-scaled mixture.

\subsubsection{M 67}

For this  cluster   \cite{teti00}   obtained [Fe/H]=-0.03$\pm$0.03  by
analysing evolved stars, including helium  core--burning stars of  the
clump.  \cite{ht91} measured [Fe/H]=-0.04$\pm$0.12. A similar value of
[Fe/H]=-0.05 was found using calibrations of the ultraviolet excess at
$(B-V)_0$=0.06  as a function   of [Fe/H] \citep{mmj93}.   \cite{fb92}
acquired spectra at a  resolution of 0.25  {\AA} for three  M~67 dwarf
stars  of  [Fe/H]  values of  -0.07,   0.05 and  0.03.   \cite{Yong05}
collected  spectra at  a resolution  of 28000  for K--giant members of
several open clusters, including  3  stars in M~67.  Iron   abundances
measured using Fe {\sc I} lines for these stars  are: 0.03, -0.05, and
-0.01 dex.   They  provided a final   result for  the cluster,  taking
account of measurements obtained using the  Fe {\sc II} lines, of 0.02
$\pm0.14$.

These results all agree reasonably with our  own.  Finally, a study on
M~67,     based also   on UVES     spectra    of main--sequence  stars
\citep{lastsofias}, measured [Fe/H]=0.03$\pm  0.03$, which agree  with
our measurement of 0.03$\pm 0.04$.

For  the remaining elements,  the comparison with \cite{lastsofias} is
also very satisfactory: the discrepancies are in most cases (aluminum,
calcium, titanium and nickel) within 0.02 dex, and only for oxygen and
sodium do the measurements  differ by as  much  as 0.08 and  0.07 dex,
respectively.  We recall that we  used the same line list but
different synthesis code and model atmosphere as \cite{lastsofias}.

The measurement  of   \cite{Yong05},  despite the  agreement   in  the
measurement   of the iron abundance,  infer  abundance ratios that are
systematically higher than ours.   The disagreement is twice  as large
as their quoted rms  errors,  or more, as in   the cases of  aluminum,
sodium and titanium.

Our conclusion is that all  studies completed are in good agreement in
their  measurement of the  M~67 iron  content, which  is equal  to the
solar value, and  most of them reach the same  conclusion also for the
abundance  of  other  elements.   However, disagreement  between  some
element abundance measurements does  exist, in particular for aluminum
and sodium.  We  note that for these two  elements the abundances rely
only on few  lines, and that the \cite{Yong05}  analysis is based only
on giant  stars.  Furthermore,  for these two  elements, discrepancies
between the abundances for dwarfs  and giants in the same cluster have
been observed \citep[see e.g.][]{lastsofias}.

\subsubsection{Praesepe}

Former studies of the metallicity    of this cluster measured   either
barely  supersolar    or definitely  supersolar.   Our  result   is in
agreement   with  those of  the  latter  group.  \cite{bb88} used high
resolution      spectra (0.1--0.2 {\AA}) of  five     F dwarfs and one
spectroscopic binary taken at the Palomar 5--meter Hale telescope, and
measured [Fe/H]=0.13$\pm$0.07.   \cite{b89}  reanalysed  three  of the
five  Praesepe F dwarf    spectra presented in  \cite{bb88}, selecting
those with low rotational velocities and well determined temperatures.
They found  [Fe/H] values of 0.033,  0.106, and 0.147, which implied a
revised  value       for    the  mean     cluster      metallicity  of
[Fe/H]=0.092$\pm$0.067.   \cite{fb92} also  measured Praesepe's   mean
metallicity and their result was [Fe/H]=0.038$\pm$0.039.  They studied
two stars in common with \cite{b89}, with which the result agreed very
well (0.033 and 0.016 dex of difference in the measurements).  In this
case a  disagreement  with our  measurement of [Fe/H]=0.27  is clearly
evident.

For this cluster,   \cite{an07} obtained from spectroscopy a  moderate
supersolar metallicity  ([Fe/H]=0.11  $\pm0.03$  dex),  which   is the
adopted value: however they  measured a higher value from  photometric
analysis  ([Fe/H]=0.20  $\pm0.04$ dex).    To  measure  both sets   of
measurements, they  adopted  published reddening   data and  performed
simultaneous parameter determination.

\begin{figure*}
\begin{center}
\resizebox{16cm}{3cm}{
\begin{tabular}{c c}
\includegraphics[width=7cm]{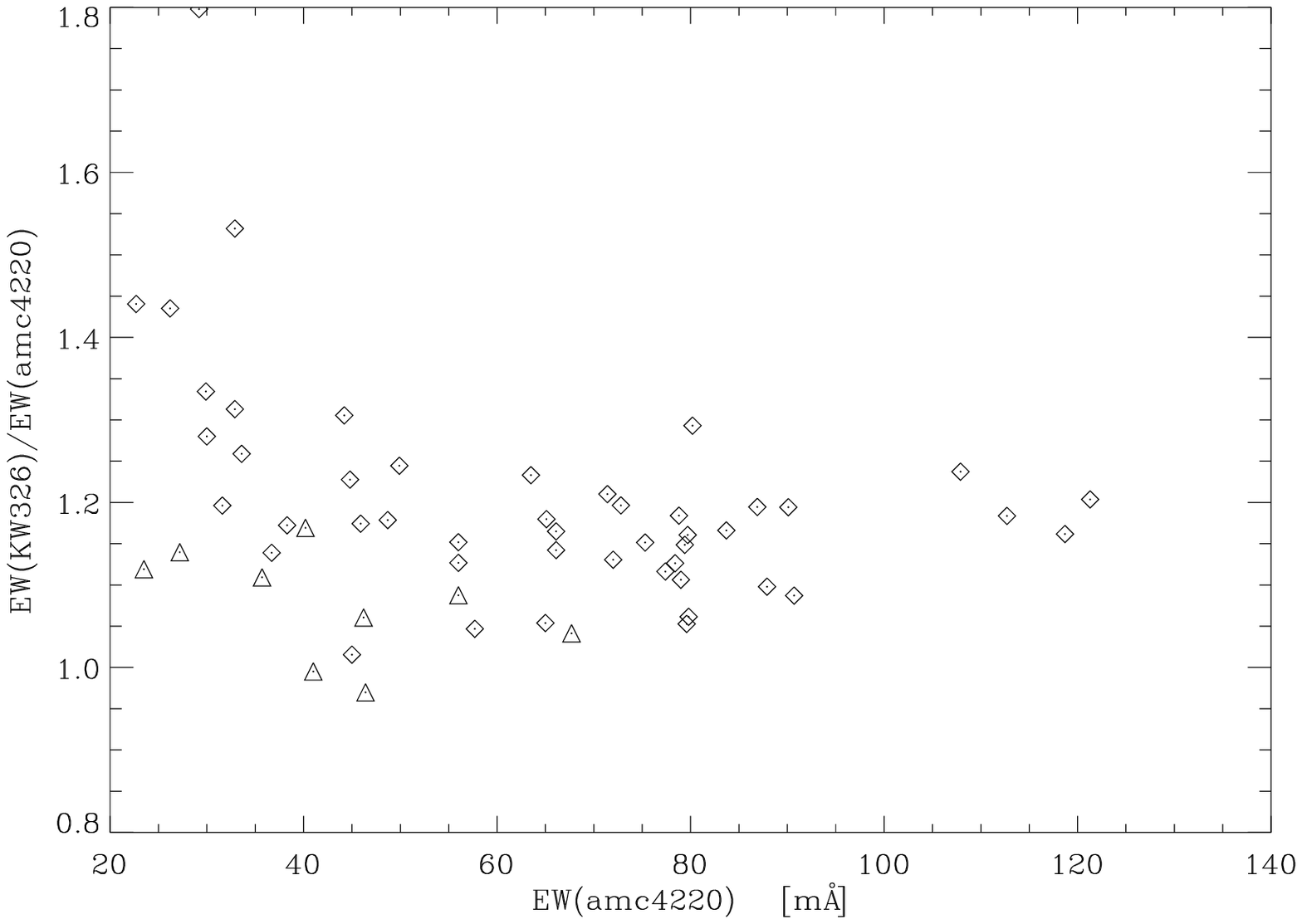}&
\includegraphics[width=7cm]{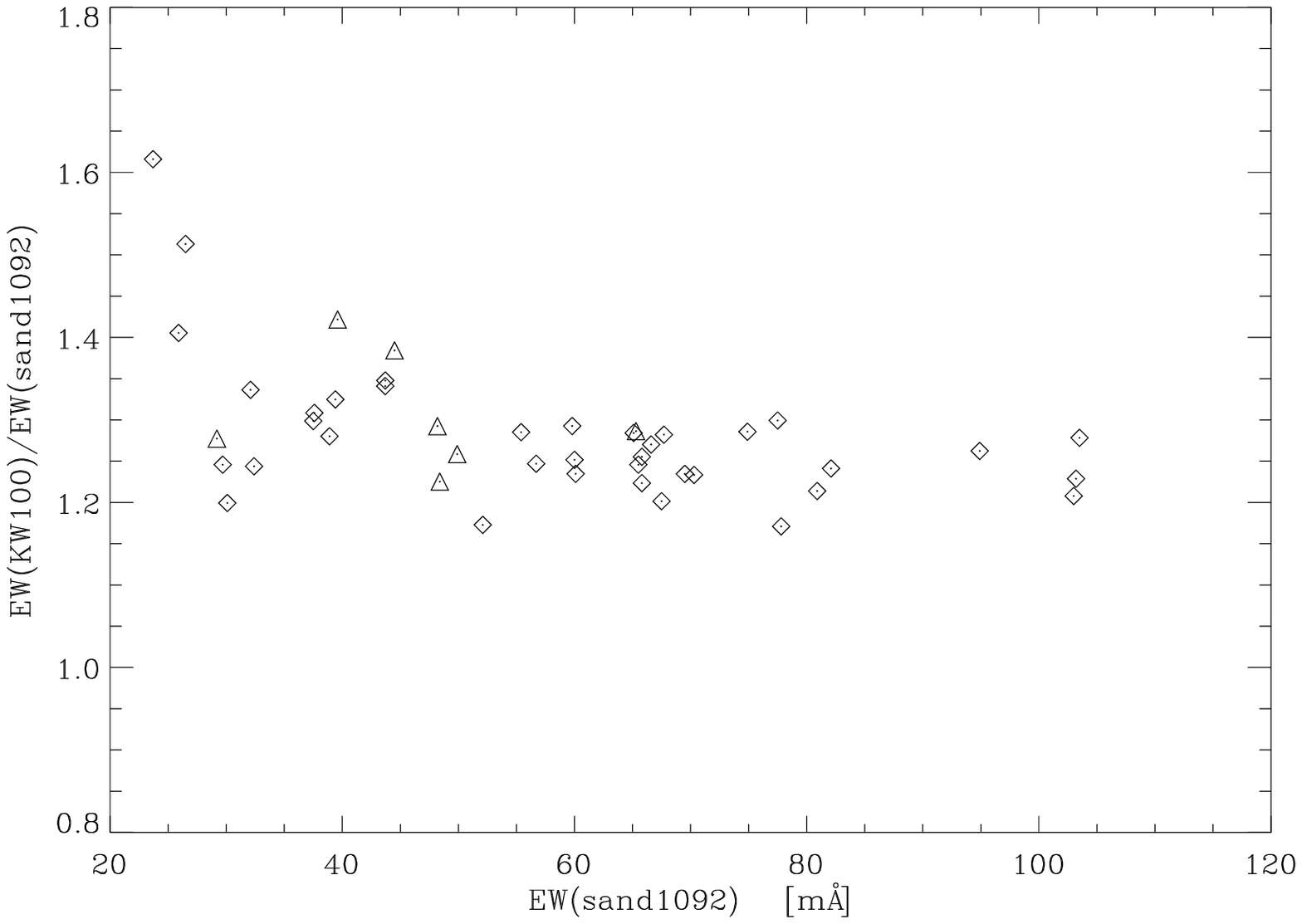}\\
\end{tabular}}
\end{center}

\caption{Comparisons between iron line  EW measurements of stars  with
different   abundances     and     with      similar   (spectroscopic)
temperatures. Two Praesepe stars are compared respectively with a star
in IC~4651  (KW~326 versus   AMC~4220, left  panel)  and in  M~67 (KW~100
versus Sanders~1092, right  panel).  Fe {\sc I}  and Fe{\sc II} lines are
plotted  with  different   symbols.    KW~100 and  Sanders~1092   have
virtually the  same  temperature. AMC~4220  is evaluated  to  be about
100~K hotter than KW~326, which reduces the  effect of the metallicity
difference on the EW ratios between the two stars.}

\label{cfrew} 
\end{figure*}

Figure \ref{cfrew} shows comparisons between iron line EW measurements
in one Praesepe  star and one star in IC~4651 (left  panel) and one in
M~67 (right panel).  For the  comparisons, we chose two pairs of stars
whose temperatures,  derived from  the spectral analysis,  differed by
only  110~K  and   10~K.   We  display  the  comparison   of  sets  of
measurements  for KW~326  against AMC~4220  (in the  left  panel), and
KW~100 against Sanders~1092 (in  the right panel of Fig. \ref{cfrew}).
The differences  between the  EWs are of  about 15$\%$ for  KW~326 and
AMC~4220 and even higher for KW~100 and Sanders~1092.

We  repeated  the  chemical  analysis  of  Praesepe  stars  using  two
different  combinations of  parameters in  addition to  those adopted.
First, we used the  values computed in Sect.  \ref{abundanceanalysis},
namely  $T_{phot}$, $G_{phot}$,  and $\xi_{fit}$.   Then, we  used the
combination of parameters that flatten the EW versus abundance and the
$\chi$  versus abundance  trends.  The  latter is  the  combination of
parameters  that we  would  have  obtained by  means  of the  spectral
analysis  if we  had ignored  photometric data.  They differ  from the
adopted  values because  the adopted  values were  adjusted  to ensure
closer  agreement  with  the  values of  $T_{phot}$,  $G_{phot}$,  and
$\xi_{fit}$, as explained in Sect.  \ref{abundanceanalysis}.
 
In both cases,  and for each  and  every star, the  abundance analysis
confirmed that  the cluster was metal--rich,  as can be seen  in Table
\ref{praesepetab}.   We  can  confidently  conclude Praesepe  shows an
enhanced  average metallicity of not   lower  than $\approx$ 0.2  dex,
which  implies that it  is closer  in metallicity to  the  Hyades than
previously understood (see Sect. \ref{discussion}).

\begin{table*}[ht] 
\begin{center}
\begin{footnotesize} 
\begin{tabular}{l c c c c c c c }
star& [Fe/H]& $\xi$&  T &  G &  ion. eq.& corr($chi$ vs.ab) %
&corr(EW vs.ab)\\
\hline
\multicolumn{8}{c}{$T=T_{phot},\ G=G_{phot}, \ \xi=\xi_{fit}$}\\
\hline
KW 49  &  0.220 & 1.15 &   6028 & 4.44 &   -0.11&    0.35 &   0.62 \\
KW 100 &  0.354 & 1.09 &   5977 & 4.45 &   -0.30&    0.10 &   0.76 \\
KW 208 &  0.202 & 1.11 &   5993 & 4.45 &   -0.22&    0.44 &   0.54 \\
KW 326 &  0.447 & 0.98 &   5873 & 4.47 &    0.07&   -0.10 &   0.54 \\
KW 368 &  0.409 & 0.91 &   5811 & 4.48 &    0.12&   -0.36 &   0.59 \\
KW 392 &  0.246 & 1.01 &   5902 & 4.46 &   -0.27&    0.39 &   0.47 \\
KW 418 &  0.187 & 1.19 &   6062 & 4.44 &   -0.16&    0.48 &   0.40 \\
\hline									 
\multicolumn{8}{c}{Minimising parameters with very loose photometric constraints}\\
\hline 	 	      	       	      	       	 	 	 	   	 	  	 
KW 49  &  0.268 & 1.58 &   6290 & 4.66 &    0.01&    0.01 &   0.01 \\
KW 100 &  0.409 & 1.76 &   6370 & 4.64 &   -0.01&   -0.22 &   0.05 \\
KW 208 &  0.322 & 1.52 &   6340 & 4.64 &    0.00&   -0.01 &  -0.02 \\
KW 326 &  0.366 & 1.14 &   5860 & 4.62 &   -0.01&   -0.06 &   0.10 \\
KW 368 &  0.166 & 1.34 &   5700 & 4.52 &    0.00&   -0.03 &   0.04 \\
KW 392 &  0.354 & 1.44 &   6250 & 4.60 &    0.00&    0.01 &   0.02 \\
KW 418 &  0.359 & 1.44 &   6450 & 4.76 &    0.00&    0.00 &   0.03 \\
\hline
\end{tabular}
\end{footnotesize} 
\end{center}
\caption
{Results of our revised abundance analyses  of Praesepe stars.  In the
upper part of the Table,  theoretical parameters are used (temperature
from the  photometry, gravity from the models  and the temperature and
microturbulence from   the fit). In the lower   part of the  Table, we
adopted the most reliable measurements of parameters found by ignoring
constraints from photometry or models. Also these analyses confirm the
high metallicity of Praesepe is evident.}
\label{praesepetab}
\end{table*}

\thispagestyle{empty}
\begin{table*}[h!]
\begin{center}
\begin{footnotesize}
\begin{tabular}{l c c c c c c c c c}
{\bf STAR}&         \multicolumn{4}{|c|}{\bf PHOTOMETRY}         &\multicolumn{2}{c|}{\bf COMPUTED   }   &\multicolumn{3}{c}{\bf SPECTROSCOPIC}\\
           &\multicolumn{3}{|c}{2MASS}&\multicolumn{1}{c|}{Johnson}   &\multicolumn{2}{c|}{\bf  VALUES}        &\multicolumn{3}{c}{\bf  VALUES }       \\
&\multicolumn{1}{|c}{J}&      H    &     K$_S$ &\multicolumn{1}{c|}{V}&$T_{eff}$ &\multicolumn{1}{c|}{ $\log G$}&$T_{eff}$&$\log G$&   $\xi$   \\
\hline
\multicolumn{10}{c}{NGC 3680}\\
\hline											                                                
AHTC 1009  &   13.094   &   12.813  &  12.730   &  14.290 & 5926 & 4.46 & 6010 & 4.50 & 1.16 \\
Eggen 70   &   13.471   &   13.157  &  13.088   &  14.589 & 6053 & 4.44 & 6210 & 4.47 & 1.36 \\
\noalign{\smallskip}						       
\hline								       
\multicolumn{10}{c}{IC 4651}\\					       
\hline    							        
AMC 1109   &  13.153    &  12.805   &  12.773   &  14.534 & 5678 & 4.50 & 6060 & 4.55 & 1.20 \\ 
AMC 2207   &  13.210    &  12.931   &  12.831   &  14.527 & 5824 & 4.48 & 6050 & 4.36 & 1.18 \\
AMC 4220   &  13.551    &  13.215   &  13.088   &  14.955 & 5605 & 4.51 & 5910 & 4.57 & 1.06 \\
AMC 4226   &  13.303    &  12.966   &  12.909   &  14.645 & 5743 & 4.49 & 5980 & 4.44 & 1.19 \\
Eggen 45   &  12.961    &  12.657   &  12.617   &  14.27  & 5844 & 4.47 & 6320 & 4.43 & 1.50 \\ 
\noalign{\smallskip}						       
\hline								       
\multicolumn{10}{c}{PRAESEPE}\\					       
\hline								        
KW  49     &   9.591    &   9.330   &   9.276   &  10.65  & 6028 & 4.44 & 6150 & 4.50 & 1.41 \\ 
KW 100     &   9.463    &   9.242   &   9.182   &  10.57  & 5977 & 4.45 & 6150 & 4.34 & 1.78 \\
KW 208     &   9.565    &   9.357   &   9.259   &  10.66  & 5993 & 4.45 & 6280 & 4.58 & 1.52 \\
KW 326     &  10.091    &   9.784   &   9.706   &  11.20  & 5873 & 4.47 & 5800 & 4.48 & 1.28 \\
KW 368     &  10.183    &   9.825   &   9.753   &  11.30  & 5811 & 4.48 & 5720 & 4.49 & 1.12 \\
KW 392     &   9.659    &   9.396   &   9.329   &  10.78  & 5902 & 4.46 & 6250 & 4.56 & 1.48 \\
KW 418     &   9.463    &   9.227   &   9.142   &  10.51  & 6062 & 4.44 & 6150 & 4.36 & 1.27 \\
\noalign{\smallskip}
\hline
\multicolumn{10}{c}{M 67}\\
\hline
Sand 746   &  13.058    &  12.746   &  12.628   &  14.380 & 5608 & 4.41 & 5750 & 4.43 & 0.84 \\
Sand 1048  &  13.189    &  12.856   &  12.804   &  14.411 & 5792 & 4.36 & 5900 & 4.37 & 0.94 \\
Sand 1092  &  13.189    &  12.856   &  12.804   &  13.308 & 5960 & 4.32 & 6160 & 4.41 & 1.42 \\
Sand 1255  &  13.216    &  12.921   &  12.844   &  14.486 & 5733 & 4.38 & 5840 & 4.48 & 1.05 \\
Sand 1283  &  12.926    &  12.630   &  12.599   &  14.115 & 5903 & 4.33 & 6100 & 4.41 & 1.16 \\
Sand 1287  &  12.835    &  12.503   &  12.442   &  14.030 & 5838 & 4.35 & 6100 & 4.41 & 1.26 \\

\hline
\end{tabular}
\end{footnotesize}
\end{center}
\caption{Comparison between  spectroscopic  and computed  values, i.e.
photometric  temperatures and gravities  computed assuming photometric
temperatures.  The photometric temperatures   are computed by means of
the infrared magnitudes from the 2MASS catalogue (shown in the second,
third and  fourth Col.), which were  converted into TCS colours, and
the available V magnitudes (shown in fifth Col.).}
\label{cfr_temp}
\end{table*}


\subsection{Comparison between photometric and spectroscopic temperatures}
\label{spph}

In  Table \ref{cfr_temp}, we  show the  photometric data  adopted, the
stellar  parameters  derived  from  these  photometric  data  and  the
spectroscopic parameters. The mean $T_{spec}-T_{phot}$ values are 120,
325,  122,  and  169  K  for NGC~3680,  IC~4651,  Praesepe,  and  M~67
respectively.   With   the  exception  of  two   Praesepe  stars,  the
photometric temperature  is always  lower than the  spectroscopic one.
Praesepe suffers  little extinction, and  its mean $T_{spec}-T_{phot}$
value, as  for NGC~3680,  is amongst the  lowest one.  However,  it is
also the  only cluster for which  the stellar values  of this quantity
are scattered  about the average value, which  is unsurprising, since,
as discussed in Sect.  \ref{paramerr}, there were more problems in the
determination of the spectroscopic parameters for Praesepe stars.  The
systematic  bias   towards  positive  $T_{spec}-T_{phot}$   has  three
possible  causes: errors  in  the colour--temperature  transformation,
errors in the adopted cluster reddening, and some systematic effect in
determining  the spectroscopic  temperature.  While  the error  in the
cluster  extinction  depends  on  the individual  cluster,  the  other
sources of systematic errors should affect all stars in a similar way.
All clusters  have a  similar average $T_{spec}-T_{phot}$,  apart from
IC~4651 for which this quantity is $\approx$ 150~K higher than in M~67
and  $\approx$ 200~K higher  than in  the other  two clusters.   It is
natural to suggest  that this is due to an  underestimation of the B-V
colour excess adopted for IC~4651  \citep{AMC1}.  If this is the case,
a value about 0.04 higher, i.e.   E(B-V) of between 0.12 and 0.13 mag,
is   the  correct   value.   Similar   conclusions  were   reached  by
\cite{przhcn} and \cite{biazzo}.

\subsection{Summary}

For most elements and all clusters, our results infer, or are at least
consistent with, solar--scaled abundances in all four target clusters,
regardless of whether  we consider $\alpha$ elements (oxygen, silicon,
calcium,   and   titanium), iron--peak elements    (iron, chromium and
nickel), or  the odd--Z elements aluminum and  sodium.  Among  the few
elements  that  are an  exception to these   trends, oxygen is 0.4 dex
below its solar--scaled value in Praesepe and 0.2  above this value for
NGC   3680.    Subsolar (slightly outside    the  margins of errors)
metallicities were also measured for aluminum in NGC 3680 and IC 4651,
and nickel in NGC 3680.

\section{The metallicity gradient and the age--metallicity relationship.} 

\label{discussion}

The   comparisons   of    our   metallicity measurements   with  other
high--quality data analyses made in  Sect.  \ref{spph}, clearly  shows
that  these studies   agree within  the quoted  errors  even when they
employ different model atmospheres and methods  of analysis.  In Table
\ref{metcomp}, we compile all the  open cluster abundance measurements
to   our  knowledge, obtained   using   high--resolution spectroscopic
analysis.  For our target clusters, abundance data and relative errors
are taken from  the present work, regardless of  whether  or not other
results are also  available.  For M~67  and IC~4651, our results  are,
however, almost indistinguishable from the mean literature value.  For
Collinder~261, Berkeley~29, and  NGC~6253, we indicate the average  of
the   two   determinations in  Table  \ref{comptab}   and half  of the
differences between  them  as abundance estimation  and relative error
respectively.    The  results  are shown  with  their  references, the
instruments  used, and the  spectral  resolution.  Cluster ages and
Galactocentric  distances are also given.   For  the latter values, in
the cases in which we used the WEBDA database, we adopted the distance
and the Galactic  coordinates from the  database, and assumed  a solar
distance  from the  Galactic centre of  8   Kpc, to be consistent  with
\cite{BOCCE}.

We considered  only    one     study   of Hyades     cluster,     i.e.
\cite{paulson03}, which was nevertheless extensive (almost one hundred
target stars) and measured [Fe/H]=0.13 $\pm$0.05  dex.  The history of
the metallicity measurements for Hyades is summarised by \cite{takeda}
(see  Fig.  8 of  this  paper):  they  range  from  $\approx 0.1$ to
$\approx 0.2$ dex, Takeda's analysis favouring the higher results.

The other element abundance ratios given in \cite{paulson03} are solar
within $1  \sigma$,   emphasising the similarity   between Hyades  and
Praesepe.

\begin{figure}
\begin{center}
\includegraphics[width=7cm]{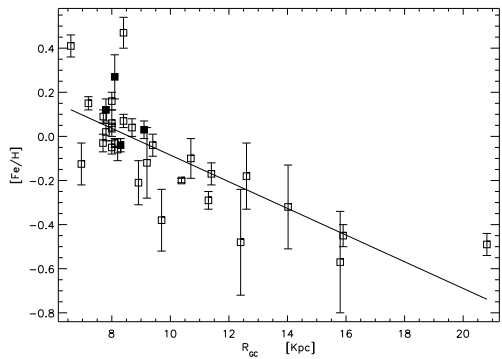}

\caption{Abundance  gradient as resulting from high--spectroscopy open
cluster chemical analyses. Data plotted and relative error bars are as
in Table \ref{metcomp}; for Hyades only,  we have adopted in this Fig.
a different value, i.e. 0.16 $\pm$ 0.04  dex, which takes into account
also other estimations  of about 0.2 dex.  The  y--axis of  the filled
symbol points refer to our datapoints.   The line represents the -0.06
dex/Kpc relation.}

\label{gradgraph} 
\end{center}
\end{figure}

The abundance  data   and   the  Galactocentric distances   in   Table
\ref{metcomp}  have  been  displayed  in  the Galactocentric--distance
versus  metallicity  diagram  of Fig.   \ref{gradgraph}.   The  Hyades
abundance used in this Fig.  was  0.16 dex, with  an error bar of 0.04
dex,  to   take into    account   both the aforementioned    result of
\cite{paulson03} (0.13 dex), and the  higher estimations at around 0.2
dex   given  by other   authors.   Error bars along   the  x--axis are
difficult to evaluate: it would indeed be  unwise to extend the limits
of the  statistical analysis too far, although  several Kpc can easily
be reached   for the most  distant  clusters. Homogeneous and reliable
photometry and open  cluster parameter determination are as  important
as  the robustness  of the  metallicity   measurements.  Of the   four
datapoints from  this  work in  Fig.  \ref{gradgraph},  only  Praesepe
departs significantly above the -0.06--dex/Kpc--steep linear fit.  The
only  other two clusters that have  an exceptionally high iron content
are the very  metal rich NGC~6253 and  NGC~6791, whose  abundances are
also solar scaled for   most of the  elements  \citep{Paola07,BOCCE3}.
Anyway,  these three clusters are  very different  from one another as
far as  age and location  are concerned.  In particular the relatively
young  and close--by Praesepe has  not formed significantly inside the
Galactic disk and drifted outwards later on, as suggested for NGC~6791
\citep{BOCCE3}.

It is  worth mentioning that the  most metal rich clusters analysed so
far, NGC~6253 and  NGC~6791, to  which  we add Praesepe,  are all very
subsolar in oxygen, which may lead us to conclude that they formed out
of  material enriched by type Ia  supernovae. However, this suggestion
is not supported by the abundance ratios  of the other alpha elements,
which are in general solar in value.

\thispagestyle{empty}

\addtocounter{table}{1}

\section{Conclusions.}

This   work is one of   several  aimed  at collecting  high--precision
abundance measurements in open clusters, based on high--resolution and
high--signal--to--noise  spectra.  We present  data for four clusters,
and confirm all of the abundance trends present  in the literature for
IC~4651  and M~67.  For these   clusters, all determinations
agree to within  a few hundredths  of dex, irrespective of whether dwarfs
or giants have been observed.  We report slightly different abundances
for NGC~3680  and mostly Praesepe, which is  shown to be substantially
metal rich.

The  modern  iron determinations  agree  very  well  in most of cases,
though there  is still no precise agreement  about the extent to which
Praesepe and Hyades are supersolar.  Metallicity measurements for open
clusters  are in general    consistent, irrespective of   the analysis
method and research group.  As for abundance  ratios of elements other
than iron,  significant  differences  between   the results  of    the
different groups still arise.

The dispersion within each cluster is limited to a few hundredths of a
dex, and differences of up to 0.1 dex or more are measured between the
individual stars.  These differences  have to be interpreted as  being
due to limitations in the observations and in the analysis rather than
true, intrinsic  chemical  variations   in  the  cluster  composition.
Caution should therefore be taken when adopting analyses that consider
only one or a few stars per cluster.

We     find  a   significant   supersolar   metallicity    for IC~4651
([Fe/H]=0.12$\pm$0.05)  and   a    solar    metallicity    for    M~67
([Fe/H]=0.03$\pm$0.04)    and   a slightly   subsolar  metallicity for
NGC~3680 ([Fe/H]=-0.04$\pm$0.03).  The  surprising  result is that  of
Praesepe,  which  is  as metal  rich   as  [Fe/H]=0.27$\pm$0.10.   The
photometric analysis  of  \cite{an07} also indicates that  Praesepe is
metal rich. The reasons for the discrepancy with other results need to
be further investigated.

For other elements, the composition is solar within the errors for all
elements and for  all the clusters,  except for  aluminum (subsolar in
NGC~3680  and   IC~4651),  nickel  (also   subsolar in NGC~3680),  and
oxygen. We   find [O/Fe] to  be  $\approx 0.2$  and  $\approx -0.4$ in
NGC~3680 and Praesepe respectively, and  just $\approx 1 \sigma$ below
the solar value in M~67.

We confirm the extraordinarily similar overall chemical composition of
the Sun and the star cluster M~67.

\begin{acknowledgements} 

  We are greatly indebted to Sofia Randich for her contribution to the
  data analysis  presented here and  for many useful  comments.  Jorge
  Melendez, Nuno Santos, Valentina D'Orazi and K.  Sinha deserve to be
  warmly thanked for their  valuable feedback. Our analysis  presented
  here  was  based on  observations  obtained  at the   ESO Very Large
  Telescope (VLT). This publication made use of data products from the
  Two  Micron   All Sky  Survey, which   is  a joint  project   of the
  University of Massachusetts and the Infrared Processing and Analysis
  Center/California  Institute of Technology,  funded by  the National
  Aeronautics  and  Space  Administration  and  the  National  Science
  Foundation; and from the WEBDA database, created by J.-C- Mermilliod
  and now operated at the institute for Astronomy of the University of
  Vienna.    The   SIMBAD  astronomical   database   and  the   NASA's
  Astrophysics Data System Abstract Service have also been extensively
  used.    G.    P.   acknowledges   the   support   of  the   project
  PTDC/CTE-AST/65971/2006  of the  Portuguese FCT  and that  of Indian
  DST, L. P. acknowledges ESO DGDF.

\end{acknowledgements}

\bibliographystyle{aa}

\longtab{3}{
\begin{footnotesize} 
\thispagestyle{empty}
\begin{longtable}{l c c c c c c c c c c c c}

\caption  []  {\label{metmeas} Table  of  the computed abundances, all
from   the neutral--element lines.  In  the  rows with the average the
standard  deviation refers to the  different stellar values within the
cluster (for NGC~3680 we give  half of the  difference between the two
stellar values). In  the other rows the  standard deviations refer  to
the measurements from  he different lines.  When no  more than 3 lines
are used for a given element, half  the difference between the largest
and the  smallest   values, rather  than   the standard deviation,  is
given. In the case of Praesepe the  errors adopted are higher than the
standard deviation: 0.10 dex.}

\\
\endfirsthead
\multicolumn{13}{l}{Table \ref{metmeas} cont.}\\
\endhead
\endfoot

STAR%
&[Fe/H]&N$_{Fe}$&$\sigma_{Fe}$%
&[Na/H]&N$_{Na}$&$\sigma_{Na}$%
&[Al/H]&N$_{Al}$&$\sigma_{Al}$%
&[Si/H]&N$_{Si}$&$\sigma_{Si}$\\
\hline	 
\noalign{\smallskip}
\multicolumn{13}{c}{{\bf {\large {NGC 3680}}}}\\
      AHTC 1009&    0.00$\pm$    0.09& 59&    0.06&    0.04$\pm$    0.06&  2&    0.00&   -0.14$\pm$    0.05&  1&    0.00&   -0.03$\pm$    0.02&  8&    0.04\\
       Eggen 70&   -0.07$\pm$    0.09& 50&    0.06&   -0.10$\pm$    0.06&  3&    0.08&   -0.11$\pm$    0.05&  1&    0.00&   -0.07$\pm$    0.03&  8&    0.05\\
                     AVERAGE&-0.04&\multicolumn{2}{c}{$\sigma$=0.03}&-0.03&\multicolumn{2}{c}{$\sigma$=0.07}&-0.12&\multicolumn{2}{c}{$\sigma$=0.02}&-0.05&\multicolumn{2}{c}{$\sigma$=0.02}\\
\hline	 
\noalign{\smallskip}
\multicolumn{13}{c}{{\bf {\large {IC 4651}}}}\\
       AMC 1109&    0.11$\pm$    0.09& 57&    0.04&    0.09$\pm$    0.06&  3&    0.05&    0.05$\pm$    0.05&  2&    0.01&    0.11$\pm$    0.02&  9&    0.04\\
       AMC 2207&    0.13$\pm$    0.09& 62&    0.06&    0.06$\pm$    0.06&  3&    0.04&    0.04$\pm$    0.05&  2&    0.02&    0.11$\pm$    0.03&  9&    0.05\\
       AMC 4220&    0.19$\pm$    0.09& 51&    0.05&    0.12$\pm$    0.07&  3&    0.07&    0.01$\pm$    0.05&  2&    0.00&    0.11$\pm$    0.02&  7&    0.03\\
       AMC 4226&    0.13$\pm$    0.09& 61&    0.09&    0.13$\pm$    0.07&  3&    0.09&    0.06$\pm$    0.05&  2&    0.03&    0.08$\pm$    0.02&  8&    0.05\\
       Eggen 45&    0.05$\pm$    0.09& 61&    0.08&    0.02$\pm$    0.06&  3&    0.06&   -0.05$\pm$    0.05&  2&    0.08&    0.09$\pm$    0.03&  8&    0.02\\
                     AVERAGE& 0.12&\multicolumn{2}{c}{$\sigma$=0.05}& 0.09&\multicolumn{2}{c}{$\sigma$=0.05}& 0.02&\multicolumn{2}{c}{$\sigma$=0.04}& 0.10&\multicolumn{2}{c}{$\sigma$=0.02}\\
\hline	 
\multicolumn{13}{c}{{\bf {\large {PRAESEPE}}}}\\
           KW49&    0.22$\pm$    0.10& 54&    0.06&    0.17$\pm$    0.06&  3&    0.05&    --             & --&  --    &    0.21$\pm$    0.03&  7&    0.06\\
          KW100&    0.27$\pm$    0.10& 39&    0.04&    0.27$\pm$    0.06&  1&    0.00&    --             & --&  --    &    0.38$\pm$    0.04&  6&    0.12\\
          KW208&    0.28$\pm$    0.10& 56&    0.05&    0.31$\pm$    0.06&  2&    0.00&    0.30$\pm$    0.05&  1&    0.00&  0.26$\pm$    0.03&  7&    0.05\\
          KW326&    0.29$\pm$    0.10& 53&    0.06&    0.23$\pm$    0.07&  3&    0.04&    0.26$\pm$    0.05&  1&    0.00&  0.29$\pm$    0.03&  7&    0.09\\
          KW368&    0.26$\pm$    0.11& 56&    0.07&    0.19$\pm$    0.07&  3&    0.05&    0.17$\pm$    0.05&  1&    0.00&  0.23$\pm$    0.03&  6&    0.07\\
          KW392&    0.35$\pm$    0.11& 55&    0.04&    0.27$\pm$    0.07&  3&    0.07&    0.26$\pm$    0.05&  2&    0.00&  0.29$\pm$    0.04&  8&    0.10\\
          KW418&    0.24$\pm$    0.10& 56&    0.07&    0.15$\pm$    0.07&  3&    0.05&    0.11$\pm$    0.05&  1&    0.00&  0.20$\pm$    0.04&  7&    0.09\\
                     AVERAGE& 0.27&\multicolumn{2}{c}{$\sigma$=0.04*}& 0.23&\multicolumn{2}{c}{$\sigma$=0.06}& 0.22&\multicolumn{2}{c}{$\sigma$=0.08}& 0.26&\multicolumn{2}{c}{$\sigma$=0.06}\\
\hline	 
\noalign{\smallskip}
\multicolumn{13}{c}{{\bf {\large {M 67}}}}\\
   Sanders  746&    0.07$\pm$    0.09& 51&    0.04&    0.04$\pm$    0.07&  3&    0.03&    0.13$\pm$    0.05&  1&    0.00&   -0.05$\pm$    0.02&  6&    0.03\\
  Sanders  1048&    0.03$\pm$    0.10& 63&    0.06&    0.00$\pm$    0.06&  3&    0.03&   -0.04$\pm$    0.05&  2&    0.02&   -0.01$\pm$    0.02&  6&    0.03\\
  Sanders  1092&    0.07$\pm$    0.09& 59&    0.04&    0.09$\pm$    0.06&  3&    0.04&    0.02$\pm$    0.05&  2&    0.00&    0.08$\pm$    0.03&  9&    0.08\\
  Sanders  1255&    0.01$\pm$    0.10& 59&    0.04&    0.00$\pm$    0.07&  3&    0.05&    --             & --&  --    &    0.02$\pm$    0.02&  7&    0.06  \\
  Sanders  1283&    0.03$\pm$    0.09& 59&    0.04&    0.04$\pm$    0.06&  3&    0.02&   -0.12$\pm$    0.05&  2&    0.00&    0.03$\pm$    0.03&  8&    0.08\\
  Sanders  1287&   -0.04$\pm$    0.09& 51&    0.03&   -0.08$\pm$    0.06&  3&    0.05&    --             & --&  --    &   -0.05$\pm$    0.03&  5&      0.04\\
                     AVERAGE& 0.03&\multicolumn{2}{c}{$\sigma$=0.04}& 0.01&\multicolumn{2}{c}{$\sigma$=0.06}& 0.00&\multicolumn{2}{c}{$\sigma$=0.10}& 0.00&\multicolumn{2}{c}{$\sigma$=0.05}\\
\hline

\newpage

STAR%
&[Ca/H]&N$_{Ca}$&$\sigma_{Ca}$%
&[Ti/H]&N$_{Ti}$&$\sigma_{Ti}$%
&[Cr/H]&N$_{Cr}$&$\sigma_{Cr}$%
&[Ni/H]&N$_{Ni}$&$\sigma_{Ni}$\\
\hline
\noalign{\smallskip}
\multicolumn{13}{c}{{\bf {\large {NGC 3680}}}}\\
      AHTC 1009&    0.05$\pm$    0.08&  9&    0.02&    0.06$\pm$    0.11&  9&    0.03&    0.00$\pm$    0.12&  5&    0.05&   -0.08$\pm$    0.06& 21&    0.05\\
       Eggen 70&   -0.05$\pm$    0.08&  9&    0.03&   -0.06$\pm$    0.09&  9&    0.06&   -0.05$\pm$    0.11&  6&    0.03&   -0.10$\pm$    0.07& 23&    0.06\\
                     AVERAGE& 0.00&\multicolumn{2}{c}{$\sigma$=0.05}& 0.00&\multicolumn{2}{c}{$\sigma$=0.06}&-0.03&\multicolumn{2}{c}{$\sigma$=0.03}&-0.09&\multicolumn{2}{c}{$\sigma$=0.01}\\

\hline
\noalign{\smallskip}
\multicolumn{13}{c}{{\bf {\large {IC 4651}}}}\\
       AMC 1109&    0.14$\pm$    0.08& 11&    0.04&    0.13$\pm$    0.11& 11&    0.05&    0.15$\pm$    0.11&  6&    0.04&    0.11$\pm$    0.06& 22&    0.05\\
       AMC 2207&    0.17$\pm$    0.08& 11&    0.05&    0.05$\pm$    0.10& 10&    0.04&    0.14$\pm$    0.11&  6&    0.05&    0.10$\pm$    0.07& 23&    0.06\\
       AMC 4220&    0.19$\pm$    0.08& 11&    0.07&    0.17$\pm$    0.11& 10&    0.08&    0.21$\pm$    0.12&  6&    0.05&    0.17$\pm$    0.06& 22&    0.07\\
       AMC 4226&    0.19$\pm$    0.08& 11&    0.08&    0.10$\pm$    0.11&  7&    0.08&    0.06$\pm$    0.11&  6&    0.08&    0.10$\pm$    0.06& 21&    0.08\\
       Eggen 45&    0.08$\pm$    0.08& 11&    0.04&    0.04$\pm$    0.09& 10&    0.06&    0.02$\pm$    0.10&  6&    0.06&    0.01$\pm$    0.07& 19&    0.08\\
                     AVERAGE& 0.16&\multicolumn{2}{c}{$\sigma$=0.04}& 0.10&\multicolumn{2}{c}{$\sigma$=0.05}& 0.12&\multicolumn{2}{c}{$\sigma$=0.07}& 0.10&\multicolumn{2}{c}{$\sigma$=0.06}\\

\hline
\multicolumn{13}{c}{{\bf {\large {PRAESEPE}}}}\\
          KW 49&    0.19$\pm$    0.09& 10&    0.05&    0.19$\pm$    0.11&  8&    0.05&    0.19$\pm$    0.13&  6&    0.04&    0.22$\pm$    0.07& 17&    0.05\\
         KW 100&    0.29$\pm$    0.10&  5&    0.07&    0.18$\pm$    0.11&  4&    0.02&    0.21$\pm$    0.12&  3&    0.04&    0.25$\pm$    0.07& 10&    0.05\\
         KW 208&    0.31$\pm$    0.09& 10&    0.10&    0.28$\pm$    0.11&  7&    0.05&    0.35$\pm$    0.12&  5&    0.03&    0.26$\pm$    0.07& 17&    0.07\\
         KW 326&    0.26$\pm$    0.10&  7&    0.02&    0.28$\pm$    0.13& 10&    0.07&    0.28$\pm$    0.14&  5&    0.05&    0.26$\pm$    0.07& 18&    0.07\\
         KW 368&    0.26$\pm$    0.10&  9&    0.05&    0.21$\pm$    0.14&  8&    0.06&    0.31$\pm$    0.14&  6&    0.01&    0.25$\pm$    0.06& 20&    0.05\\
         KW 392&    0.35$\pm$    0.09& 10&    0.07&    0.34$\pm$    0.11&  9&    0.06&    0.38$\pm$    0.13&  6&    0.05&    0.33$\pm$    0.08& 23&    0.05\\
         KW 418&    0.25$\pm$    0.09& 10&    0.08&    0.13$\pm$    0.11&  7&    0.03&    0.26$\pm$    0.13&  5&    0.03&    0.20$\pm$    0.08& 20&    0.04\\
                     AVERAGE& 0.27&\multicolumn{2}{c}{$\sigma$=0.05}& 0.23&\multicolumn{2}{c}{$\sigma$=0.07}& 0.28&\multicolumn{2}{c}{$\sigma$=0.07}& 0.25&\multicolumn{2}{c}{$\sigma$=0.04}\\

\hline
\noalign{\smallskip}
\multicolumn{13}{c}{{\bf {\large {M 67}}}}\\
   Sanders  746&    0.12$\pm$    0.08& 11&    0.05&    0.06$\pm$    0.12& 11&    0.05&    0.15$\pm$    0.12&  6&    0.09&    0.00$\pm$    0.05& 15&    0.03\\
  Sanders  1048&    0.09$\pm$    0.08& 10&    0.05&    0.04$\pm$    0.11&  9&    0.03&    0.11$\pm$    0.12&  6&    0.04&    0.05$\pm$    0.06& 21&    0.06\\
  Sanders  1092&    0.11$\pm$    0.08& 11&    0.04&    0.13$\pm$    0.10& 11&    0.06&    0.08$\pm$    0.11&  6&    0.06&    0.07$\pm$    0.07& 20&    0.06\\
  Sanders  1255&    0.03$\pm$    0.08& 10&    0.05&   -0.01$\pm$    0.11& 10&    0.06&    0.04$\pm$    0.12&  6&    0.06&    0.04$\pm$    0.06& 23&    0.06\\
  Sanders  1283&    0.06$\pm$    0.08& 11&    0.06&    0.00$\pm$    0.10& 10&    0.05&    0.02$\pm$    0.11&  5&    0.04&    0.02$\pm$    0.07& 22&    0.04\\
  Sanders  1287&   -0.05$\pm$    0.08& 11&    0.03&   -0.16$\pm$    0.10&  8&    0.03&   -0.07$\pm$    0.11&  5&    0.07&   -0.10$\pm$    0.07& 22&    0.06\\
                     AVERAGE& 0.06&\multicolumn{2}{c}{$\sigma$=0.06}& 0.01&\multicolumn{2}{c}{$\sigma$=0.10}& 0.06&\multicolumn{2}{c}{$\sigma$=0.08}& 0.01&\multicolumn{2}{c}{$\sigma$=0.06}\\
\hline

\end{longtable}
\end{footnotesize}
}


\longtab{6}{
\begin{landscape}
\begin{table*}[ht]
\begin{center}
\begin{tabular}{c c c c c c c c c c c c}

\hline
[Fe/H]        &[O/Fe]        &[Al/Fe]       &[Ni/Fe]       &[Na/Fe]       &[Si/Fe]       &[Ca/Fe]       &[Ti/Fe]       &N&Reference &R       &Instr/OBS   \\
\hline
\multicolumn{12}{c}{IC 4651}\\
\hline
 0.12$\pm$0.05&     --       &-0.10$\pm$0.06&-0.02$\pm$0.08&-0.03$\pm$0.07&-0.02$\pm$0.05& 0.04$\pm$0.06&-0.02$\pm$0.07&4&1         &100K    &UVES/VLT    \\
 0.11$\pm$0.01&     --       &     --       &     --       &     --       &     --       &     --       &     --       &5&4         &48K     &FEROS/1.5m  \\
 0.10$\pm$0.03&     --       & 0.01$\pm$0.07& 0.05$\pm$0.05& 0.02$\pm$0.16& 0.07$\pm$0.03& 0.02$\pm$0.04& 0.09$\pm$0.04&22&10       &100K    &UVES/VLT    \\
\multicolumn{12}{c}{ }\\
\hline
\multicolumn{12}{c}{Praesepe}\\
\hline
 0.27$\pm$0.10&-0.4 $\pm$0.2 &-0.05$\pm$0.12&-0.02$\pm$0.10&-0.04$\pm$0.12&-0.01$\pm$0.12& 0.00$\pm$0.11&-0.04$\pm$0.12&7&1         &100K    &UVES/VLT    \\
 0.04$\pm$0.04&     --       &     --       &     --       &     --       &     --       &     --       &     --       &7&7         &60\&30K &CFHT \& Hale\\
 0.11$\pm$0.03&     --       &     --       &     --       &     --       &     --       &     --       &     --       &4&2         &55K     &MIKE/MagCl  \\
\multicolumn{12}{c}{ }\\
\hline
\multicolumn{12}{c}{M 67}\\
\hline
 0.03$\pm$0.04&-0.07$\pm$0.08&-0.03$\pm$0.11&-0.02$\pm$0.07&-0.02$\pm$0.07&-0.03$\pm$0.06& 0.03$\pm$0.07&-0.02$\pm$0.11&6&1         &100K    &UVES/VLT    \\
 0.02$\pm$0.14& 0.07$\pm$0.03& 0.17$\pm$0.05& 0.08$\pm$0.10& 0.30$\pm$0.10& 0.09$\pm$0.11& 0.07$\pm$0.06& 0.12$\pm$0.07&3&14        &28K     &KPNO \& CTIO\\
 0.03$\pm$0.03& 0.01$\pm$0.03&-0.05$\pm$0.04&-0.02$\pm$0.04& 0.05$\pm$0.07& 0.02$\pm$0.04& 0.05$\pm$0.04&-0.02$\pm$0.04&10&11       &45K     &UVES/VLT    \\
-0.03$\pm$0.03& 0.02$\pm$0.04& 0.14$\pm$0.06& 0.04$\pm$0.06& 0.19$\pm$0.07& 0.10$\pm$0.04& 0.04$\pm$0.08& 0.04$\pm$0.10&12&13       &60\&30K &SOFIN/NOT   \\
 0.02$\pm$0.12&     --       &     --       &     --       &     --       &     --       &     --       &     --       &3& 7        &60K     &CFHT        \\
-0.04$\pm$0.12&     --       &     --       &     --       &     --       &     --       &     --       &     --       &8& 9        &34K     &KPNO        \\
\multicolumn{12}{c}{ }\\
\hline
\multicolumn{12}{c}{Collinder 261}\\
\hline
-0.03$\pm$0.03&-0.2 $\pm$0.1 & 0.12$\pm$0.08& 0.06$\pm$0.08& 0.33$\pm$0.06& 0.24$\pm$0.05& 0.01$\pm$0.05&-0.12$\pm$0.09&6&5         &48K     &FEROS/1.5m  \\
-0.22$\pm$0.11&-0.1 $\pm$0.15& 0.39$\pm$0.12& 0.02$\pm$0.04& 0.48$\pm$0.22& 0.22$\pm$0.09&-0.04$\pm$0.10&-0.07$\pm$0.09&7&8         &25K     &CTIO        \\
\multicolumn{12}{c}{ }\\
\hline
\multicolumn{12}{c}{Be 29}\\
\hline
-0.54$\pm$0.02& 0.23$\pm$0.03& 0.26$\pm$0.01&-0.02$\pm$0.02& 0.36$\pm$0.01& 0.18$\pm$0.02& 0.02$\pm$0.02& 0.33$\pm$0.04&2&14        &28K     &KPNO \& CTIO\\
-0.44$\pm$0.18& 0.18$\pm$0.02& 0.20$\pm$0.03& 0.11$\pm$0.06& 0.39$\pm$0.08& 0.22$\pm$0.03& 0.10$\pm$0.06& 0.02$\pm$0.01&2&3         &34K     &HIRES/KECK  \\
\multicolumn{12}{c}{ }\\
\hline
\multicolumn{12}{c}{NGC 6253}\\
\hline
 0.46$\pm$0.03&-0.18$\pm$0.06&-0.08$\pm$0.12&-0.05$\pm$0.01& 0.21$\pm$0.02& 0.09$\pm$0.06&-0.02$\pm$0.12&-0.19$\pm$0.10&4&6         &43K     &UVES/VLT    \\
 0.36$\pm$0.07&   --         &    --        & 0.08$\pm$0.07&     --       & 0.02$\pm$0.08&-0.04$\pm$0.12&-0.01$\pm$0.14&5&12        &47K     &UVES/VLT    \\
\hline
\end{tabular}
\end{center}
\caption
{Literature  data  about   high--resolution spectroscopic   studies of
clusters that have more  than one high--resolution entry. Errors refer
to  the rms  dispersion  between stellar values, or   half of the full
spread when   only two or  three  stars are  studied.  From Col.  1 to
Col. 8 we indicate the abundance measurements.   In Col. 9, the number
of  stars used is given;  in Col.  10, the  reference, in  Col. 11 and
Col.  12, we indicate, respectively,  the  spectral resolution and the
instrument and/or the observatory.}
\label{comptab} 
\end{table*}
References
(1)Present study;
(2)\cite{an07};        
(3)\cite{Giovanni04};  
(4)\cite{cbgt04};      
(5)\cite{Carretta05};  
(6)\cite{BOCCE3};      
(7)\cite{fb92};        
(8)\cite{frieletal03}; 
(9)\cite{ht91};        
(10)\cite{przhcn};      
(11)\cite{lastsofias};  
(12)\cite{Paola07};     
(13)\cite{teti00};      
(14)\cite{Yong05}.
\end{landscape}
}

\longtab{9}{  
\begin{longtable}{l c c c c c c c c}  
\caption[]
{\label{metcomp} Compilation   of  high--resolution  studies  on  open
clusters.    The first  panel  shows  the   results of  the  abundance
analyses,  while  the  second  one  summarises  dataset  and clusters'
properties, showing, in the given order, the number of stars used, the
reference (i.e. number  in the bibliography  and "PPF" for the present
study;   some of  the cited      references for  the abundances    are
compilations of the results obtained  by the same group), the spectral
resolution  power,  the instrument  and/or   the telescope   employed,
clusters' Galactocentric distances and  ages.   The two latter  values
are flagged  with a letter  according  to the   reference used: F  for
\cite{fj93}, B for \cite{BOCCE}, C for \cite{cc94} and W for the WEBDA
database.}
\\ 
\hline 
\endfirsthead
\multicolumn{9}{l}{Table \ref{metcomp} cont.}\\ 
\endhead 
\hline
Cluster        &[Fe/H]        &[O/Fe]        &[Al/Fe]       &[Ni/Fe]       &[Na/Fe]       &[Si/Fe]       &[Ca/Fe]       &[Ti/Fe]       \\
\hline									                                                                 
$\alpha$  Per  &-0.06$\pm$0.05&     --       &     --       &     --       &     --       &     --       &     --       &     --       \\
Be 17          &-0.10$\pm$0.09& 0.00$\pm$0.05& 0.25$\pm$0.09& 0.02$\pm$0.09& 0.37$\pm$0.08& 0.30$\pm$0.05&-0.04$\pm$0.03&-0.1 $\pm$0.08\\
Be 20          &-0.45$\pm$0.05& 0.18$\pm$0.05& 0.18$\pm$0.01&-0.02$\pm$0.02& 0.2 $\pm$0.1 & 0.05$\pm$0.03& 0.08$\pm$0.01& 0.38$\pm$0.06\\
Be 22          &-0.32$\pm$0.19&     --       & 0.28$\pm$0.05& 0.04$\pm$0.01& 0.04$\pm$0.05&-0.04$\pm$0.1 &-0.08$\pm$0.02& 0.11$\pm$0.1 \\
Be 29          &-0.49$\pm$0.05& 0.21$\pm$0.02& 0.23$\pm$0.03& 0.05$\pm$0.06& 0.38$\pm$0.01& 0.20$\pm$0.02& 0.06$\pm$0.04& 0.18$\pm$0.16\\
Be 31          &-0.57$\pm$0.23& 0.24$\pm$0.08& 0.22$\pm$0.13& 0.11$\pm$0.12& 0.27$\pm$0.10& 0.20$\pm$0.14& 0.13$\pm$0.05& 0.08$\pm$0.09\\
Be 32          &-0.29$\pm$0.04&     --       & 0.11$\pm$0.10& 0.00$\pm$0.04& 0.12$\pm$0.07& 0.12$\pm$0.04& 0.07$\pm$0.04& 0.02$\pm$0.06\\
Be 66          &-0.48$\pm$0.24&     --       & 0.00$\pm$0.2 & 0.24$\pm$0.25& 0.15$\pm$0.2 &     --       &-0.05$\pm$0.2 & 0.43$\pm$0.2 \\
Blanco 1       & 0.04$\pm$0.02& 0.02$\pm$0.11&     --       &-0.18$\pm$0.01&     --       &-0.09$\pm$0.02&-0.09$\pm$0.02&-0.10$\pm$0.03\\
Collinder 261  &-0.12$\pm$0.09&-0.15$\pm$0.05& 0.25$\pm$0.13& 0.04$\pm$0.02& 0.41$\pm$0.07& 0.23$\pm$0.01&-0.01$\pm$0.02&-0.09$\pm$0.02\\
Coma  Ber      &-0.05$\pm$0.03&     --       &     --       &     --       &     --       &     --       &     --       &     --       \\
Hyades         & 0.13$\pm$0.05&     --       &     --       &     --       & 0.01$\pm$0.09& 0.05$\pm$0.05& 0.07$\pm$0.07& 0.03$\pm$0.05\\ 
IC 2391        & 0.06$\pm$0.06&     --       &     --       &     --       &     --       &     --       &     --       &     --       \\
IC 4651        & 0.12$\pm$0.05&     --       &-0.10$\pm$0.06&-0.02$\pm$0.08&-0.03$\pm$0.07&-0.02$\pm$0.05& 0.04$\pm$0.06&-0.02$\pm$0.07\\
IC 4665        &-0.03$\pm$0.04&     --       &     --       & 0.05$\pm$0.13&     --       & 0.09$\pm$0.19& 0.03$\pm$0.14& 0.21$\pm$0.17\\
M 35           &-0.21$\pm$0.10&     --       &     --       &     --       &     --       &     --       &     --       &     --       \\
M 67           & 0.03$\pm$0.04&-0.07$\pm$0.08&-0.03$\pm$0.11&-0.02$\pm$0.07&-0.02$\pm$0.07&-0.03$\pm$0.06& 0.03$\pm$0.07&-0.02$\pm$0.11\\
NGC 188        &-0.12$\pm$0.16&     --       &     --       &     --       &     --       &     --       &     --       &     --       \\
NGC 2141       &-0.18$\pm$0.15& 0.00$\pm$0.06& 0.18$\pm$0.07& 0.04$\pm$0.11& 0.41$\pm$0.04& 0.05$\pm$0.19& 0.10$\pm$0.04& 0.24$\pm$0.11\\
NGC 2324       &-0.17$\pm$0.05&     --       & 0.00$\pm$0.08&-0.09$\pm$0.02& 0.31$\pm$0.10& 0.06$\pm$0.11& 0.15$\pm$0.05&-0.08$\pm$0.03\\
NGC 2477       & 0.07$\pm$0.03&     --       &-0.01$\pm$0.04& 0.00$\pm$0.04& 0.12$\pm$0.03& 0.05$\pm$0.03&-0.01$\pm$0.01& 0.01$\pm$0.06\\
NGC 2506       &-0.20$\pm$0.01&     --       &     --       &     --       &     --       &     --       &     --       &     --       \\
NGC 2660       & 0.04$\pm$0.04&     --       &-0.08$\pm$0.10&-0.03$\pm$0.02& 0.12$\pm$0.04& 0.00$\pm$0.03& 0.04$\pm$0.05& 0.00$\pm$0.03\\
NGC 3680       &-0.04$\pm$0.03& 0.2 $\pm$0.05&-0.08$\pm$0.04&-0.05$\pm$0.03& 0.01$\pm$0.08&-0.01$\pm$0.04& 0.04$\pm$0.06& 0.04$\pm$0.07\\
NGC 3960       & 0.02$\pm$0.04&     --       &-0.06$\pm$0.06&-0.01$\pm$0.03& 0.09$\pm$0.03& 0.04$\pm$0.05& 0.02$\pm$0.03&-0.04$\pm$0.02\\
NGC 6134       & 0.15$\pm$0.03&     --       &     --       &     --       &     --       &     --       &     --       &     --       \\
NGC 6253       & 0.41$\pm$0.05&-0.18$\pm$0.06&-0.08$\pm$0.12&-0.01$\pm$0.06& 0.21$\pm$0.02& 0.05$\pm$0.03&-0.03$\pm$0.01&-0.1 $\pm$0.1 \\
NGC 6791       & 0.47$\pm$0.07&-0.31$\pm$0.08&-0.21$\pm$0.09&-0.07$\pm$0.07& 0.13$\pm$0.21&-0.01$\pm$0.10&-0.15$\pm$0.08& 0.03$\pm$0.09\\
NGC 6819       & 0.09$\pm$0.03&     --       &-0.07$\pm$0.07& 0.01$\pm$0.02& 0.47$\pm$0.07& 0.18$\pm$0.04&-0.04$\pm$0.06&-0.01$\pm$0.04\\
NGC 7789       &-0.04$\pm$0.05&-0.07$\pm$0.09& 0.18$\pm$0.08&-0.02$\pm$0.05& 0.28$\pm$0.07& 0.14$\pm$0.05&     --       &-0.03$\pm$0.07\\
Pleiades       &-0.03$\pm$0.02&     --       &     --       &     --       &     --       &     --       &     --       &     --       \\
Praesepe       & 0.27$\pm$0.10&-0.4 $\pm$0.2 &-0.05$\pm$0.12&-0.02$\pm$0.1 &-0.04$\pm$0.12&-0.01$\pm$0.12& 0.00$\pm$0.11&-0.04$\pm$0.12\\
Saurer 1       &-0.38$\pm$0.14& 0.47$\pm$0.13& 0.33$\pm$0.02& 0.20$\pm$0.05& 0.44$\pm$0.05& 0.38$\pm$0.1 & 0.20$\pm$0.04& 0.12$\pm$0.12\\
\hline
\newpage
\hline
Cluster        &N&Reference           &R &\multicolumn{2}{c}{Instr/OBS}      & R$_{GC}$ [Kpc]&Age [Gyr]   &\\
\hline								   
$\alpha$  Per  &6&  9            &60\&30K &\multicolumn{2}{c}{CFHT \& Hale            }&  8.2  W& 0.07W   &\\
Be 17          &3& 10                 &25K &\multicolumn{2}{c}{KPNO                   }& 10.7  W&12   W   &\\
Be 20          &2& 18                 &28K &\multicolumn{2}{c}{KPNO \& CTIO           }& 15.9  W& 6.0 W   &\\
Be 22          &2& 17                 &34K &\multicolumn{2}{c}{HIRES/KECK             }& 14.02 B& 2.40B   &\\
Be 29          &\multicolumn{5}{c}{See Table \ref{comptab}}                            & 20.81 B& 3.70B   &\\
Be 31          &1& 18                 &28K &\multicolumn{2}{c}{KPNO \& CTIO           }& 15.8  W& 2.0 W   &\\
Be 32          &9& 4,13               &47K &\multicolumn{2}{c}{UVES/VLT               }& 11.30 B& 6.5 B   &\\
Be 66          &1& 17                 &34K &\multicolumn{2}{c}{HIRES/KECK             }& 12.4  W& 5.0 W   &\\
Blanco 1       &8&  8                 &50K &\multicolumn{2}{c}{Anglo/Aus              }&  8.0  W& 0.62W   &\\
Collinder 261  &\multicolumn{5}{c}{See Table \ref{comptab}}                            &  6.96 B& 6   B   &\\
Coma  Ber      &14& 9                 &60K &\multicolumn{2}{c}{CFHT                   }&  8.0  W& 0.4 W   &\\
Hyades         &98&12                 &60K &\multicolumn{2}{c}{HIRES/KECK             }&  8.0  W& 0.8 W   &\\
IC 2391        &66&14                 &50K &\multicolumn{2}{c}{FEROS/2.2m MPG--ESO    }&  8.0  W& 0.05 W  &\\
IC 4651        &4& 1                  &100K&\multicolumn{2}{c}{UVES/VLT               }&  7.8  L& 1.6 C   &\\
IC 4665        &18& 15                &60K &\multicolumn{2}{c}{HIRES/KECK             }&  7.7  W& 0.43W   &\\
M 35           &9& 2                  &20K &\multicolumn{2}{c}{WIYN/HYDRA             }&  8.91 B& 0.18B   &\\
M 67           &6& 1                  &100K&\multicolumn{2}{c}{UVES/VLT               }&  9.1  F& 4.8 C   &\\
NGC 188        &7& 11                 &34K &\multicolumn{2}{c}{KPNO                   }&  9.2  W& 4.3 W   &\\
NGC 2141       &1& 18                 &28K &\multicolumn{2}{c}{KPNO \& CTIO           }& 12.6  F& 4   F   &\\
NGC 2324       &7& 4                  &47K &\multicolumn{2}{c}{UVES/VLT               }& 11.4  W& 0.4 W   &\\
NGC 2477       &6& 4                  &47K &\multicolumn{2}{c}{UVES/VLT               }&  8.4  W& 0.7 W   &\\
NGC 2506       &2& 6                  &48K &\multicolumn{2}{c}{FEROS/1.5m             }& 10.38 B& 1.70B   &\\
NGC 2660       &5& 4,13               &47K &\multicolumn{2}{c}{UVES/VLT               }&  8.69 B& 0.95B   &\\
NGC 3680       &2& 1                  &100K&\multicolumn{2}{c}{UVES/VLT               }&  8.3  F& 1.8 C   &\\
NGC 3960       &6& 4,13               &47K &\multicolumn{2}{c}{UVES/VLT               }&  7.80 B& 1.2 B   &\\
NGC 6134       &6& 6              &48K\&43K&\multicolumn{2}{c}{FEROS/1.5m \& UVES/VLT }&  7.2  W& 0.93W   &\\
NGC 6253       &\multicolumn{5}{c}{See Table \ref{comptab}}                            &  6.6  B& 3   B   &\\  
NGC 6791       &5& 7                  &43K &\multicolumn{2}{c}{UVES/VLT               }&  8.4  F& 4.4 W   &\\
NGC 6819       &3& 3                  &40K &\multicolumn{2}{c}{SARG/TNG               }&  7.71 B& 2   B   &\\
NGC 7789       &9& 16                 &30K &\multicolumn{2}{c}{SOFI/NOT               }&  9.4  F& 2   F   &\\
Pleiades       &13& 9             &60\&30K &\multicolumn{2}{c}{CFHT \& Hale           }&  8.1  W& 0.1 W   &\\
Praesepe       &7&  1                 &100K&\multicolumn{2}{c}{UVES/VLT               }&  8.1  W& 0.7 W   &\\
Saurer 1       &2& 5                  &34K &\multicolumn{2}{c}{HIRES/KECK             }&  9.7  W& 7.1 W   &\\
\hline
\end{longtable}
References.
(1)Present study;
(2)\cite{BYN};         
(3)\cite{Bragaglia01}; 
(4)\cite{BOCCE4};      
(5)\cite{Giovanni04};  
(6)\cite{cbgt04};      
(7)\cite{BOCCE3};      
(8)\cite{Ford05};      
(9)\cite{fb92};        
(10)\cite{frieletal05}; 
(11)\cite{ht90};        
(12)\cite{paulson03};   
(13)\cite{Paola06};     
(14)\cite{ppm}
(15)\cite{Shen05};      
(16)\cite{teti05};      
(17)\cite{Sandro05};    
(18)\cite{Yong05}.
}
\end{document}